\documentclass[final,5p,times]{elsarticle}
\usepackage{graphicx,amsmath,mathtools}
\usepackage{hyperref}
\usepackage{bm,makecell}
\usepackage{siunitx}
\usepackage{color}
\usepackage{bold-extra}
\usepackage{multirow}
\usepackage[T1]{fontenc}
\usepackage{braket}
\usepackage{bm}
\usepackage{arydshln}

\newcommand{\codeversion}{4.0}
\newcommand{\hfbtho}{{\sc hfbtho}}
\newcommand{\hfodd}{{\sc hfodd}}

\newcommand{\di}{i}

\definecolor{db}{rgb}{0.180,0.543,0.340}

\newcounter{leteq}

\newcommand{\hodd}{\bf\color{red}}
\newcommand{\htho}{\bf\color{blue}}
\newcommand{\spc}{{\ }}
\newcommand{\pr}[1]{{\sc{\lowercase{#1}}}}

\newcommand{\gras}[1]{\boldsymbol{#1}}
\newcommand{\tensor}[1]{\mathsf{#1}}

\newcommand{\keyw}{{\bf Keyword:}}
\newcommand{\key}[1]{\vspace{1ex}\noindent\keyw{\spc}{\tk{#1}}
                         \newline\phantom{\keyw{\spc}{\tk{#1}}}{\spc}}

\newcommand{\be}{\begin{equation}}
\newcommand{\ee}{\end{equation}}
\newcommand{\ba}{\begin{array}}
\newcommand{\ea}{\end{array}}
\newcommand{\bn}{\begin{eqnarray}}
\newcommand{\en}{\end{eqnarray}}
\newcommand{\bc}{\begin{center}}
\newcommand{\ec}{\end{center}}
\newcommand{\bi}{\begin{itemize}}
\newcommand{\ei}{\end{itemize}}

\newcommand{\tk}[1]{{\tt{#1}}\index{#1}}

\renewcommand{\tk}[1]{\textcolor{blue}   {{\tt{#1}}}}

\renewcommand{\tk}[1]{{\tt{#1}}}

\newcounter{bla}

\allowdisplaybreaks

\journal{Computer Physics Communications}

\begin{document}

\begin{frontmatter}

\title{Axially-deformed solution of the Skyrme-Hartree-Fock-Bogoliubov 
equations \newline using the transformed harmonic oscillator basis (IV) \textsc{hfbtho} 
(v{\codeversion}): \newline A new version of the program}

\author[a,b]{P. Marevi\'c}
\author[a]{N. Schunck\corref{author}}
\author[c]{E. M. Ney}
\author[d]{R. Navarro P\'erez}
\author[a]{M. Verriere}
\author[e]{J. O'Neal}

\cortext[author] {Corresponding author.\\\textit{E-mail address:} schunck1@llnl.gov}
\address[a]{Nuclear and Chemical Science Division, Lawrence Livermore National Laboratory, Livermore, CA 94551, USA}
\address[b]{Department of Physics, Faculty of Science, University of Zagreb, HR-10000 Zagreb, Croatia}
\address[c]{Department of Physics and Astronomy, CB 3255, University of North Carolina, Chapel Hill, North Carolina 27599-3255, USA}
\address[d]{Department of Physics, San Diego State University, 5500 Campanile Drive, San Diego, California 02182-1233, USA}
\address[e]{Mathematics and Computer Science Division, Argonne National Laboratory, Lemont, IL 60439, USA}

\begin{abstract}
We describe the new version {\codeversion} of the code \pr{hfbtho} that solves
the nuclear Hartree-Fock-Bogoliubov problem by using the deformed harmonic oscillator basis in cylindrical coordinates. In the new 
version, we have implemented the restoration of rotational, particle 
number, and reflection symmetry for even-even nuclei. The restoration of rotational symmetry
does not require using bases closed under rotation. Furthermore, we added the SeaLL1 functional and
improved the calculation of the Coulomb potential.
Finally, we refactored the code to facilitate
maintenance and future developments.
\end{abstract}

\begin{keyword}
Nuclear many-body problem; Density functional theory; Energy density functional 
theory; Self-consistent mean field; Hartree-Fock-Bogoliubov theory; 
Finite-temperature Hartree-Fock-Bogoliubov theory; Skyrme interaction; Gogny 
force; Pairing correlations; Pairing regularization; Collective inertia; Harmonic oscillator; 
Transformed harmonic oscillator; Restoration of symmetries;
Angular momentum projection; Particle number 
projection; Shared memory parallelism; Distributed memory 
parallelism.
\end{keyword}

\end{frontmatter}

{\bf \noindent PROGRAM SUMMARY}
\vspace{1mm}

\begin{small}
\noindent
{\em Program title:} \pr{hfbtho} v{\codeversion} \\
{\em CPC Library link to program files:} * \\
{\em Licensing provisions:} GPLv3  \\
{\em Programming language:} Fortran 2003 \\
{\em Journal reference of previous version:} 
R. N. P\'erez, N. Schunck, R.-D. Lasseri, C. Zhang and J. Sarich, Comput. Phys. Commun. 
220 (2017) 363 \\
{\em Does the new version supersede the previous version:} Yes   \\
{\em Reasons for the new version:} This version adds new capabilities to restore broken symmetries and determine corresponding
quantum numbers of even-even nuclei \\
{\em Summary of revisions:}
\begin{enumerate}
\item Angular momentum projection for even-even nuclei in a deformed basis;
\item Particle number projection for even-even nuclei in the quasiparticle basis;
\item Implementation of the SeaLL1 functional;
\item Expansion of the Coulomb potential onto Gaussians;
\item MPI-parallelization of a single {\hfbtho} execution;
\item Code refactoring.
\end{enumerate}
{\em Nature of problem:}
{\hfbtho} is a physics computer code that is used to model the structure of the
nucleus. It is an implementation of the energy density functional (EDF) 
approach to atomic nuclei, where the energy of the nucleus is obtained by 
integration over space of some phenomenological energy density, which is itself 
a functional of the neutron and proton intrinsic densities. In the present 
version of {\hfbtho}, the energy density is derived either from the zero-range 
Skyrme or the finite-range Gogny effective two-body interaction between 
nucleons. Nuclear superfluidity is treated at the Hartree-Fock-Bogoliubov (HFB) 
approximation. Constraints on the nuclear shape allow probing the potential 
energy surface of the nucleus as needed, e.g., for the description of shape 
isomers or fission. A local scale transformation of the single-particle basis 
in which the HFB solutions are expanded provides a tool to properly compute the 
structure of weakly-bound nuclei. Restoration of the rotational, particle 
number, and reflection symmetry for even-even nuclei enables recovering the quantum numbers that are 
lost at the HFB approximation.\\
{\em Solution method:}
The program uses the axial harmonic oscillator (HO) or the transformed harmonic 
oscillator (THO) single-particle basis to expand quasiparticle wave functions. 
It iteratively diagonalizes the HFB Hamiltonian based on generalized 
Skyrme-like energy densities and zero-range pairing interactions or the 
finite-range Gogny force until a self-consistent solution is found. Lagrange 
parameters are used to impose constraints on HFB solutions, and their value
is updated at each iteration from an approximation of the quasiparticle random
phase approximation (QRPA) matrix. Symmetry 
restoration is implemented through standard projection techniques.
Previous versions of the program were presented in [1-3]. \\
{\em Additional comments including restrictions and unusual features:}\\
Axial and time-reversal symmetries are assumed in HFB calculations; 
$y$-simplex symmetry and even particle numbers are assumed in angular momentum 
projection.\\

\end{small}


\section{Introduction}
\label{sec:intro}

Over the past decades, the nuclear energy density
functional (EDF) framework has become a tool of
choice for describing the properties of nuclear
structure and reactions across the entire nuclide chart
\cite{schunck2019energy,bender2003selfconsistent,
niksic2011relativistic,robledo2019mean}.
It closely resembles density functional theory
(DFT), a method widely used in condensed matter
physics and quantum chemistry, insofar that
it employs the mean-field approximation to map
a complex many-body problem onto a computationally feasible
one-body problem. 
In nuclear physics, the EDF
framework is typically realized at two distinct levels. 
The single-reference energy density 
functional (SR-EDF) method 
introduces relatively simple functionals
of nucleon
densities and currents, describing the
nuclear ground states in terms of
symmetry-breaking mean-field wave functions.
Most of the EDF-based computer programs available 
on the market correspond to different flavors of 
the SR-EDF method; see, 
e.g., \cite{perez2017axially,schunck2017solution,ryssens2015solution,
niksic2014dirhb,carlsson2010solution,bennaceur2005coordinatespace} for some 
selected examples. 
However, a more advanced description
requires the inclusion of collective correlations
related to the restoration of broken symmetries
and quantum shape fluctuations.
This is the basic tenet of the multi-reference 
energy density functional (MR-EDF) method.

The previous versions of the {\hfbtho} program are largely implementations of 
the SR-EDF formalism in the axial harmonic oscillator (HO) basis 
or the transformed harmonic oscillator (THO) basis \cite{stoitsov2005axially,
stoitsov2013axially,perez2017axially}. The core of the program is a solver for the 
self-consistent Hartree-Fock-Bogoliubov (HFB) equation. While the initial 
release \cite{stoitsov2005axially}
was restricted to even-even nuclei with Skyrme EDFs and 
contact pairing interactions, more recent versions expanded the theoretical 
framework significantly: to describe parity-breaking shapes, nuclei with odd number of 
particles, and nuclei at finite temperature \cite{stoitsov2013axially}; to solve the 
HFB equation for the finite-range Gogny potentials, compute the 
collective mass tensor and zero-point energy corrections, regularize the 
pairing interaction, and compute properties of fission fragments
\cite{perez2017axially}. 

Among the publicly available codes, MR-EDF capabilities include  
the restoration of particle number symmetry in the canonical basis in 
{\hfbtho} (all versions) and the restoration
of rotational, isospin, particle-number, and reflection symmetries of HFB states in 
{\hfodd} 3.06h \cite{dobaczewski2021solution}. Note that {\hfodd} projects either on total 
particle number $A$ or total isospin projection $T_z$ but not separately on the 
number of protons $Z$ and neutrons $N$.
Compared to previous versions of {\hfbtho}, the present release contains a much more expanded MR-EDF toolkit for symmetry restoration that 
is tailored for large-scale applications of the MR-EDF framework. 
Specifically, the version \codeversion~of {\hfbtho} implements
the restoration of rotational, particle number,
and reflection symmetry for even-even nuclei. These restorations
can be performed either independently 
(e.g., either the rotational and reflection symmetries only or the particle number symmetry only), 
or they can be combined in the joint 
restoration of all three types of quantum numbers (angular momentum, particle number, and parity). 
In addition, our implementation of the
angular momentum restoration bypasses the need to use
rotationally-invariant, closed bases. Symmetry restoration can now be performed in the
deformed (stretched) HO basis typically employed in large-scale calculations of 
potential energy surfaces.

In Section \ref{sec:modifs}, we review the modifications introduced in this 
version of the program.
In Section \ref{sec:benchmarks}, we give several numerical 
benchmarks for the new capabilities. Finally, in Section \ref{sec:input}, 
we discuss the new options available in the input file and explain how to 
run the code.


\section{Modifications introduced in version {\codeversion}}
\label{sec:modifs}

In this section, we present the new features added to the code between
version 3.00 and \codeversion. 


\subsection{Restoration of Broken Symmetries}
\label{subsec:amp}

A module for restoration of broken symmetries 
is the main new feature of version \codeversion.
In the following, we describe the 
underlying
theoretical framework in detail.


\subsubsection{General Framework}
\label{subsubsec:general}

The HFB states break several symmetries of the nuclear Hamiltonian and 
consequently do not carry the associated good quantum numbers. Since its first 
published version, the \pr{hfbtho} program has implemented the particle number 
restoration in the canonical basis for even-even nuclei.
The current version includes a new module
for the simultaneous restoration of 
rotational, particle number, and reflection symmetry of the HFB states for even-even nuclei
\cite{schunck2019energy,sheikh2021symmetry,
bally2021projection}. 

The main ingredient of symmetry-restoring calculations are kernels of the 
form
\begin{equation}
\mathcal{O}_{\bm{q} \bm{q}}^{J MK; NZ; p} = \braket{\Phi_{\bm{q}}
| \hat{O} \hat{P}^J_{MK} \hat{P}^N \hat{P}^Z \hat{P}^p| \Phi_{\bm{q}}}.
\label{eq:kernel}
\end{equation}
Here, $\ket{\Phi_{\bm{q}}}$ is an HFB state at point $\bm{q}$ in the collective 
space defined by the set of active constraints on the HFB solution, while 
$\hat{O}$ is either the identity operator for the norm overlap kernel, 
$\mathcal{O}_{\bm{q} \bm{q}}^{J MK; NZ; p}
\equiv \mathcal{N}_{\bm{q} \bm{q}}^{J MK; NZ; p}$, or the Hamiltonian operator 
for the Hamiltonian kernel, $\mathcal{O}_{\bm{q} \bm{q}}^{J MK; NZ; p} \equiv 
\mathcal{H}_{\bm{q} \bm{q}}^{J MK; NZ; p}$. 

The operator that projects an HFB state onto a state with good values of 
angular momentum $J$ reads
\begin{equation}
\hat{P}^J_{MK} = \frac{2J+1}{16 \pi^2} \int d\Omega\; 
D^{J*}_{MK}(\alpha, \beta, \gamma)
\hat{R}(\alpha,\beta,\gamma),
\end{equation}
where $\alpha$, $\beta$, and $\gamma$ are the usual Euler angles, 
$\int \,d \Omega \equiv \int_0^{2\pi} \,d\alpha \int_0^{\pi} 
\,d\beta \sin\beta  \int_0^{4 \pi} \,d\gamma$, and 
$D^{J}_{MK}(\alpha, \beta, \gamma)$ is the Wigner $D$-matrix 
\cite{varshalovich1988}. The coordinate-space rotation operator reads
\begin{equation}
\hat{R}(\alpha, \beta, \gamma) = e^{-\di \alpha \hat{J}_z}
e^{-\di \beta \hat{J}_y} e^{-\di \gamma \hat{J}_z}.
\end{equation}
Note that the conservation of number parity 
\cite{ring2004} allows reducing the integration
interval 
over $\gamma$ to $[0,2\pi]$. This has no practical consequence in 
{\hfbtho} since integrals over Euler angles $\alpha$ and $\gamma$ are trivial 
and can be carried out analytically
due to the axial 
symmetry. In addition, the current version of {\hfbtho} computes kernels (\ref{eq:kernel}) for the identity
and the Hamiltonian operator only. For such scalar operators,
only the $M=K=0$ components of the 
total angular momentum do not vanish identically. 

Furthermore, the operator that projects an HFB state onto a state with a good 
number of particles reads
\begin{equation}
\hat{P}^{X} = \frac{1}{2\pi} \int_0^{2\pi}d\varphi \, e^{\di (\hat{X}-X_0)\varphi},
\end{equation}
where $X = N\,(Z)$ is a label referring to neutrons (protons), $X_0 = N_0\,(Z_0)$ is the desired number 
of neutrons (protons), and $\hat{X} = \hat{N}\,(\hat{Z})$ is the neutron (proton) number 
operator. In practice, the integration interval 
over the gauge angle $\varphi$ can be reduced 
to $[0, \pi]$ using the property of a good number parity of an HFB state. The resulting integral is further discretized and particle 
number projection is performed using the Fomenko expansion 
\cite{fomenko1970projection}
\begin{equation}
\hat{P}^{X} = \frac{1}{N_{\varphi}} \sum_{l_{\tau}=1}^{N_{\varphi}}
e^{\di (\hat{X}- X_0) \varphi_{l_\tau}}, \quad \varphi_{l_{\tau}} =
\frac{\pi}{N_{\varphi}}l_{\tau},
\end{equation}
where $\tau = n\,(p)$ for neutrons (protons) and
$N_{\varphi}$ is the corresponding number of gauge angle points which may in principle
be different for neutrons and protons.

Finally, the operator that projects an HFB state onto a state with good parity 
reads
\begin{equation}
\hat{P}^{p} = \frac{1}{2} \Big(1 + p \hat{\Pi} \Big),
\end{equation}
where $p = +1\,(-1)$ for positive (negative) parity and $\hat{\Pi}$ is the 
standard parity operator \cite{egido1991parityprojected}. 

Combining the expressions for projection operators and assuming the same number of gauge angle points for neutrons and protons, the kernels 
\eqref{eq:kernel} can be written as
\begin{align}
\begin{split}
\mathcal{O}_{\bm{q} \bm{q}}^{J; NZ; p} &= \frac{2J+1}{2} 
\int_0^{\pi} d\beta\, \sin\beta\, d^{J*}_{00}(\beta) \\ & \times
\frac{1}{N_{\varphi}^2} \sum_{l_n=1}^{N_{\varphi}} 
\sum_{l_p=1}^{N_{\varphi}} e^{-\di N_0 \varphi_{l_n}} 
e^{-\di Z_0 \varphi_{l_p}} \\& \times \frac{1}{2} 
\Big[ \mathcal{O}_{\bm{q} \bm{q}}(\beta,\varphi_{l_n}, \varphi_{l_p})
+ p \mathcal{O}_{\bm{q} \bm{q}}^{\pi}(\beta,\varphi_{l_n}, \varphi_{l_p})
\Big],
\end{split}
\label{eq:kernels}
\end{align}
with the rotated kernels
\begin{subequations}
\begin{align}
\mathcal{O}_{\bm{q} \bm{q}}(\beta,\varphi_{l_n}, 
\varphi_{l_p}) &\equiv 
\braket{\Phi_{\bm{q}}|\hat{O}e^{-\di \beta \hat{J}_y} 
e^{\di \varphi_{l_n} \hat{N}}
e^{\di \varphi_{l_p} \hat{Z}} |\Phi_{\bm{q}}}, 
\label{eq:overlap_a}
\\ 
\mathcal{O}_{\bm{q} \bm{q}}^{\Pi}(\beta,\varphi_{l_n}, 
\varphi_{l_p}) & \equiv \braket{\Phi_{\bm{q}}|\hat{O}e^{-\di \beta 
\hat{J}_y} e^{\di \varphi_{l_n} \hat{N}}  
e^{\di \varphi_{l_p} \hat{Z}} \hat{\Pi} |\Phi_{\bm{q}}}.
\label{eq:overlap_b}
\end{align}
\end{subequations}
The expression for kernels can be further simplified by using the symmetries of an
HFB state. In particular,
the anti-linear $y$-time-simplex operator $\hat{S}_y^T = \hat{\Pi} \hat{T} e^{-\di \pi \hat{J}_y}$ fixes a phase through a symmetry transformation \cite{dobaczewski2000point,
dobaczewski2000pointa,bally2021projection}
\begin{equation}
\hat{S}_y^T \ket{\Phi_{\bm{q}}} = \ket{\Phi_{\bm{q}}}.
\end{equation}
Using the time-reversal symmetry, we then obtain
the following relation for the rotated kernels
\begin{align}
\mathcal{O}_{\bm{q} \bm{q}}^{\Pi}(\beta,\varphi_{l_n}, \varphi_{l_p}) =
\mathcal{O}_{\bm{q} \bm{q}}(\pi - \beta,\varphi_{l_n}, \varphi_{l_p}).
\end{align}
This greatly facilitates calculations because only the rotated kernels 
$\mathcal{O}_{\bm{q} \bm{q}}(\beta,\varphi_{l_n}, \varphi_{l_p})$ need to be 
evaluated explicitly. Moreover, since only
diagonal kernels are considered in this version of the code,
the second subscript $\bm{q}$ can
be dropped. Therefore,
the rotated kernels will simply be 
denoted as $\mathcal{O}_{\bm{q}}(\beta,\varphi_{l_n}, \varphi_{l_p})$.

The symmetry-restoring framework enables us to expand an HFB state 
$\ket{\Phi_{\bm{q}}}$ into a basis of states with good quantum numbers (angular momentum, particle number, parity) and to extract their respective coefficients \cite{ring2004}. For example, in the case of the particle number decomposition, we can
write
\begin{equation}
\ket{\Phi_{\bm{q}}} = \sum_{N} \sum_{Z} c_{\bm{q}}^{NZ}\ket{NZ},
\end{equation}
and the coefficients satisfy
\begin{equation}
\big| c_{\bm{q}}^{NZ} \big|^2 = \frac{1}{N_{\varphi}^2} \sum_{l_n=1}^{N_{\varphi}} 
\sum_{l_p=1}^{N_{\varphi}} e^{-\di N_0 \varphi_{l_n}} 
e^{-\di Z_0 \varphi_{l_p}} \mathcal{O}_{\bm{q}} (0, \varphi_{l_n}, 
\varphi_{l_p} ),
\label{eq:decomposition_NZ}
\end{equation}
with $\sum_{N} \sum_{Z} |c_{\bm{q}}^{NZ}|^2\!=\! 1$. 
Similarly, a decomposition onto states with
good angular momenta and parity
implies that the coefficients satisfy
\begin{align}
\begin{split}
\big| c_{\bm{q}}^{J;p} \big|^2 &= \frac{2J + 1}{2} 
\int_0^{\pi} 
d\beta\, \sin\beta\, d^{J*}_{00}(\beta) \\ &  \times \frac{1}{2} 
\Big[ \mathcal{O}_{\bm{q}}(\beta, 0, 0)
+ p \mathcal{O}_{\bm{q}}(\pi - \beta, 0, 0)
\Big],
\label{eq:decomposition_J}
\end{split}
\end{align}
with $\sum_J \sum_p |c_{\bm{q}}^{J;p}|^2 \!= \! 1$. Note that only collective states obeying the natural 
spin-parity selection rule, $p=(-1)^J$, are accessible within the present 
model. The coefficients of the simultaneous expansion onto states with good 
angular momentum, particle number, and parity are given by 
Eq.~\eqref{eq:kernels}, i.e., $|c_{\bm{q}}^{J;NZ;p}|^2 = 
\mathcal{O}_{\bm{q} \bm{q}}^{J; NZ; p}$. They satisfy the sum rule 
$\sum_J \sum_p \sum_{N,Z} |c_{\bm{q}}^{J;NZ;p}|^2  = 1$. Finally, the energy of a 
symmetry-restored state is calculated as
\begin{equation}
E_{\bm{q}}^{J; NZ; p}=\frac{\mathcal{H}_{\bm{q} 
}^{J; NZ; p}}{\mathcal{N}_{\bm{q}}^{J; NZ; p}}.
\label{eq:projected_energy}
\end{equation} 


\subsubsection{Bases Not Closed Under Rotation}
\label{subsubsec:bases}

Numerous implementations
of the symmetry-restoring framework
(see Refs.
\cite{niksic2011relativistic,robledo2019mean,egido2016stateoftheart} and references therein for 
some recent results)
relied on the expansion of HFB states in 
spherical HO bases
that are closed under rotation. However, such 
an approach becomes computationally intractable when describing extremely heavy 
or deformed configurations like those appearing in studies of nuclear 
fission or the structure of superheavy nuclei. In these cases, 
numerical convergence can typically
be achieved only by expanding HFB states in deformed HO bases
with incomplete oscillator shells.
However, such bases are not closed under 
rotation and the conventional symmetry-restoring framework is consequently 
inapplicable\footnote{Alternatively, symmetry restoration can also be performed with 
HFB states obtained in a coordinate-space representation \cite{bender2003selfconsistent}. To avoid the 
large computational cost associated to spatial rotations of
HFB states during the
angular momentum projection, the relevant kernels are often computed in the 
canonical basis. This can lead to similar difficulties as using incomplete HO bases; 
see \cite{avez2012evaluation,valor2000configuration,baye1984angular} for a 
discussion.}.

The elegant solution to this hurdle was proposed almost three decades ago by L. Robledo
\cite{robledo1994practical}, who reformulated Wick's theorem 
\cite{balian1969nonunitary,hara1979quantum} to encompass bases not closed under 
rotation. The first implementations of the modified symmetry-restoring 
framework were reported only very recently 
\cite{marevic2020fission,marevic2021angular}. Version {\codeversion} of 
\pr{hfbtho} is the first one to contain this capability. In particular, for 
the case of bases not closed under rotation, the rotated norm overlap kernel 
for particle type $\tau = n, p$ reads
\begin{equation}
\mathcal{N}^{(\tau)}_{\bm{q}} (\bm{x}^{(\tau)}) = 
\sqrt{\det \big[ A_{\bm{q}}^{(\tau)}(\bm{x^{(\tau)}}) \big] 
\det \big[ R(\bm{x^{(\tau)}}) \big]},
\label{eq:overlap}
\end{equation}
where $\bm{x}^{(\tau)} \equiv \{ \beta, \varphi_{l_{\tau}} \}$, $R(\bm{x}^{(\tau)})$
is the total rotation matrix, and the $A_{\bm{q}}^{(\tau)}(\bm{x}^{(\tau)})$ 
matrix reads
\begin{equation}
A_{\bm{q}}^{(\tau)} (\bm{x}^{(\tau)}) 
= 
U_{\bm{q}}^{(\tau) T} \big[ R^T(\bm{x}^{(\tau)}) \big]^{-1} U_{\bm{q}}^{(\tau) *} 
\!+ 
V_{\bm{q}}^{(\tau) T} R (\bm{x}^{(\tau)}) V_{\bm{q}}^{(\tau) *}.
\label{eq:amatrix}
\end{equation}
Here, the Bogoliubov matrices $U_{\bm{q}}^{(\tau)}$, $V_{\bm{q}}^{(\tau)}$ 
correspond to the HFB solution $\ket{\Phi_{\bm{q}}}$ for particle $\tau$. Without breaking the isospin symmetry, the full rotated norm overlap kernel is separable in isospin
\begin{equation}
\mathcal{N}_{\bm{q}}(\beta,\varphi_{l_n},\varphi_{l_p})=
\mathcal{N}_{\bm{q}}^{(\tau = n)}(\beta, \varphi_{l_n}) 
\times  
\mathcal{N}_{\bm{q}}^{(\tau = p)}(\beta, \varphi_{l_p}).
\end{equation}  
Moreover, in the case of a basis closed under rotation
we have 
$|\det[R(\bm{x}^{(\tau)})]|=1$, and the expression \eqref{eq:overlap} reduces to 
the conventional Onishi formula \cite{onishi1966generator}. 

Furthermore, the rotated density and pairing tensors for particle type $\tau$ read
\begin{subequations}
\begin{align}
\rho^{(\tau)}_{\bm{q}}(\bm{x}^{\tau}) &=
R ({\bm{x}^{(\tau)}})
V_{\bm{q}}^{(\tau) *} \Big[A^{(\tau)}_{\bm{q}}(\bm{x}^{(\tau)})\Big]^{-1}
V_{\bm{q}}^{(\tau) T} ,
\label{eq:rotated_density} \\
\kappa^{(\tau)}_{\bm{q}}(\bm{x}^{(\tau)}) &= 
R(\bm{x}^{(\tau)}) V_{\bm{q}}^{(\tau) *} 
\Big[A^{(\tau)}_{\bm{q}}(\bm{x}^{(\tau)})\Big]^{-1} U_{\bm{q}}^{(\tau) T},
\label{eq:rotated_pairing_tensor} \\
\kappa^{* (\tau)}_{\bm{q}}(\bm{x}^{(\tau)}) &= 
-R^*(\bm{x}^{(\tau)}) U_{\bm{q}}^{(\tau) *} 
\Big[A^{(\tau)}_{\bm{q}}(\bm{x}^{(\tau)})\Big]^{-1} V_{\bm{q}}^{(\tau) T}.
\label{eq:rotated_pairing_tensor*}
\end{align}
\end{subequations}
The rotated Hamiltonian kernel $\mathcal{H}_{ 
\bm{q}}(\beta,\varphi_{l_n}, \varphi_{l_p})$ is a functional of the rotated 
density and pairing tensors; see 
Section \ref{subsubsec:hamiltonian_kernel}
and Refs. \cite{schunck2019energy,
bender2003selfconsistent}
for more details.


\subsubsection{Structure of Matrices in the $y$-Simplex Basis}
\label{subsubsec:matrices}

The rotation by an angle $\beta$ about the $y$-axis of the reference frame breaks 
the axial symmetry of HFB solutions. Computations can thus 
be facilitated by using a non-axially-symmetric, computationally-efficient representation 
of the Bogoliubov matrices $U_{\bm{q}}^{(\tau)}$ and $V_{\bm{q}}^{(\tau)}$. This is 
achieved by introducing the $y$-simplex basis.

\paragraph{The $y$-simplex Basis}
The HO basis states $\ket{\alpha}$ are characterized by the set of 
quantum numbers $\{ \alpha \} = \{ n_z^{\alpha}, n_{\perp}^{\alpha}, 
\Lambda^{\!\alpha}, \Sigma^{\alpha} \}$, where $n_z^{\alpha}$ and 
$n_{\perp}^{\alpha}$ represent the number of quanta (nodes) in the $z-$ and the 
$r_{\perp}-$ direction, respectively, while $\Lambda^{\!\alpha}$ and 
$\Sigma^{\alpha} (\equiv \ket{\uparrow}, \ket{\downarrow})$ denote the 
components of the orbital angular momentum and of the spin along the $z-$axis. 
Starting from these initial basis states, it is straightforward to show that 
the linear combinations
\begin{align}
\begin{split}
\ket{n_z^{\alpha} n_{\perp}^{\alpha} \Lambda^{\!\alpha}; +} 
&= \frac{1}{\sqrt{2}} \Big[i \ket{n_z^{\alpha} n_{\perp}^{\alpha}
\Lambda^{\! \alpha} \! \uparrow } + \ket{n_z^{\alpha} n_{\perp}^{\alpha}
\!-\!\Lambda^{\! \alpha} \! \downarrow }   \Big], \\ 
\ket{n_z^{\alpha} n_{\perp}^{\alpha} \Lambda^{\! \alpha}; -} 
&= \frac{1}{\sqrt{2}} \Big[ \ket{n_z^{\alpha} n_{\perp}^{\alpha} 
\Lambda^{\! \alpha} \! \uparrow } + i \ket{n_z^{\alpha} 
n_{\perp}^{\alpha}\! - \!\Lambda^{\! \alpha}\! \downarrow }   \Big],
\end{split}
\label{eq:simplexbasis}
\end{align}
are eigenstates of the $y$-simplex operator $\hat{R}_y$ with eigenvalues of 
$+\di$ and $-\di$, respectively. The $y$-simplex operator 
$\hat{R}_y$ is defined as a rotation around the $y$-axis by an angle $\pi$, 
followed by the parity transformation $\hat{\Pi}$
\begin{equation}
\hat{R}_y = \hat{\Pi} \exp(-i \pi \hat{J}_y).\\
\label{eq:simplexoperator}
\end{equation}
The $y$-simplex basis can be used to reduce the computational cost 
by exploiting symmetries of the problem at hand.

\paragraph{Bogoliubov Matrices} 
In the $y$-simplex basis, the Bogoliubov matrices acquire the block 
structure
\begin{equation}
 U^{(\tau)}_{\bm{q}} = \Bigg( \begin{array}{cc}
        u^{(\tau)}_{\bm{q}} & 0 \\
        0 & u_{\bm{q}}^{(\tau) *} 
        \end{array} \Bigg), \quad \quad 
 V^{(\tau)}_{\bm{q}} = \Bigg( \begin{array}{cc}
        0 & -v_{\bm{q}}^{(\tau)*} \\
        v^{(\tau)}_{\bm{q}} & 0 
        \end{array} \Bigg).
\label{eq:uvsimplex}
\end{equation}
In this expression, the basis states are organized in two blocks: the first 
block comprises all states with an eigenvalue $+\di$, while the second block 
comprises all states with an eigenvalue $-\di$. The transformation between the 
components $k$ of Bogoliubov matrices in the $y$-simplex basis and the HO basis 
reads
\begin{subequations}
\begin{align}
u_{\bm{q}, k}^{(\tau) [n_z^{\alpha}, n_{\perp}^{\alpha}, \Omega^{\alpha}-\frac{1}{2}]} 
& = 
(+1) U_{\bm{q}, k}^{(\tau) [n_z^{\alpha}, n_{\perp}^{\alpha}, \Omega^{\alpha}-\frac{1}{2}, \Sigma^{\alpha}=+\frac{1}{2}]},  
\\ 
u_{\bm{q}, k}^{(\tau) [n_z^{\alpha}, n_{\perp}^{\alpha}, -\Omega^{\alpha}-\frac{1}{2}]} 
&= 
(+\di) U_{\bm{q}, k}^{(\tau) [n_z^{\alpha}, n_{\perp}^{\alpha}, \Omega^{\alpha}+\frac{1}{2}, \Sigma^{\alpha}=-\frac{1}{2}]}, 
\\
v_{\bm{q}, k}^{(\tau) [n_z^{\alpha}, n_{\perp}^{\alpha}, \Omega^{\alpha}-\frac{1}{2}]} 
&= 
(-1) V_{\bm{q}, k}^{(\tau) [n_z^{\alpha}, n_{\perp}^{\alpha}, \Omega^{\alpha}-\frac{1}{2}, \Sigma^{\alpha}=+\frac{1}{2}]},  
\\ 
v_{\bm{q}, k}^{(\tau) [n_z^{\alpha}, n_{\perp}^{\alpha}, -\Omega^{\alpha}-\frac{1}{2}]} 
&= 
(-\di) V_{\bm{q}, k}^{(\tau) [n_z^{\alpha}, n_{\perp}^{\alpha}, \Omega^{\alpha}+\frac{1}{2}, \Sigma^{\alpha}=-\frac{1}{2}]}.
\end{align}
\end{subequations}
Using these expressions, one can construct $U^{(\tau)}_{\bm{q}}$ and 
$V^{(\tau)}_{\bm{q}}$ matrices in the $y$-simplex basis from the HFB solutions 
expressed in the HO basis.

\paragraph{Rotation Matrix}
The total rotation operator corresponds to
the combination of a spatial rotation for an 
angle $\beta$ and a gauge space rotation for an angle $\varphi_{l_{\tau}}$. In 
the $y$-simplex basis, the rotation matrix acquires the following block structure
\begin{equation}
R(\bm{x}^{(\tau)}) = e^{\di \varphi_{l_{\tau}}} 
\Bigg( 
\begin{array}{cc}
        r(\beta) & 0 \\
        0 & r^*(\beta) 
\end{array} 
\Bigg),
\end{equation}
where the matrix elements $r_{\alpha \gamma}(\beta)$ of the $r(\beta)$ matrix 
read
\begin{align}
\begin{split}
r_{\alpha \gamma}(\beta) &= \frac{1}{2}\cos\Big(\frac{\beta}{2}\Big) 
\braket{n_z^{\alpha} n_{\perp}^{\alpha} 
\Lambda^{\!\alpha}|e^{-\di \beta \hat{L}_y}  
| n_z^{\gamma} n_{\perp}^{\gamma} \Lambda^{\!\gamma}} \\  & +\frac{1}{2}\cos\Big(\frac{\beta}{2}\Big)  \braket{n_z^{\alpha} 
n_{\perp}^{\alpha} \! - \!\Lambda^{\! \alpha}| e^{-\di \beta \hat{L}_y} | 
n_z^{\gamma} n_{\perp}^{\gamma}\! -\!\Lambda^{\!\gamma}}  \\ & + 
\frac{\di}{2}\sin\Big(\frac{\beta}{2}\Big) \braket{n_z^{\alpha} 
n_{\perp}^{\alpha} \Lambda^{\! \alpha}|e^{-\di \beta \hat{L}_y}  | 
n_z^{\gamma} n_{\perp}^{\gamma}\! -\!\Lambda^{\! \gamma}} \\ &+ \frac{\di}{2}\sin\Big(\frac{\beta}{2}\Big) \braket{n_z^{\alpha} 
n_{\perp}^{\alpha} \! - \!\Lambda^{\! \alpha}| e^{-\di \beta \hat{L}_y} | 
n_z^{\gamma} n_{\perp}^{\gamma} \Lambda^{\! \gamma}}.
\end{split}
\label{eq:rotation_matrix}
\end{align}
Matrix elements of the $e^{-\di \beta \hat{L}_y}$ operator are evaluated using
the prescription of Ref. \cite{nazmitdinov1996representation}.

\paragraph{Calculation of Overlaps}
Using the block structure of the Bogoliubov matrices and of the total rotation 
matrix, we can recast the $A^{(\tau)}_{\bm{q}}(\bm{x}^{(\tau)})$ matrix in 
the $y$-simplex basis as
\begin{equation}
A_{\bm{q}}^{(\tau)}(\bm{x}^{(\tau)}) = 
\Bigg( \begin{array}{cc}
        {a_{\bm{q}}^{(\tau) ++}}(\bm{x}^{(\tau)}) & 0 \\
        0 & {a_{\bm{q}}^{(\tau) --}}(\bm{x}^{(\tau)}) 
\end{array} \Bigg),
\label{eq:arecast}
\end{equation}
where 
\begin{subequations}
\begin{align}
{a_{\bm{q}}^{(\tau) ++}}(\bm{x}^{(\tau)}) 
&= 
e^{-\di \varphi_{l_{\tau}}} a_{U_{\bm{q}}}^{(\tau)}(\beta) 
+ 
e^{\di \varphi_{l_{\tau}}} a_{V_{\bm{q}}}^{(\tau)}(\beta), 
\\
{a_{\bm{q}}^{(\tau) --}}(\bm{x}^{(\tau)}) 
&= 
e^{-\di \varphi_{l_{\tau}}} \Big[ a_{U_{\bm{q}}}^{(\tau)}(\beta) \Big]^*
+ 
e^{\di \varphi_{l_{\tau}}}  \Big[ a_{V_{\bm{q}}}^{(\tau)}(\beta) \Big]^*, 
\end{align}
\end{subequations}
and
\begin{subequations}
\begin{align}
a_{U_{\bm{q}}}^{(\tau)}(\beta) 
&= \big[ u_{\bm{q}}^{(\tau)} \big]^{T} \big[ r^T(\beta) \big]^{-1} u_{\bm{q}}^{(\tau) *}, 
\\
a_{V_{\bm{q}}}^{(\tau)}(\beta) 
&= \big[v_{\bm{q}}^{(\tau)}\big]^{T} r^*(\beta) v_{\bm{q}}^{(\tau) *}.
\end{align}
\end{subequations}
The rotated norm overlap kernel then reads
\begin{equation}
\mathcal{N}_{\bm{q}}^{(\tau)}(\bm{x}^{(\tau)})  = 
  \sqrt{\det
  \Bigg[ \Bigg( \begin{array}{cc}
    {n_{\bm{q} }^{(\tau) ++}}(\bm{x}^{(\tau)}) 
        & 0 \\
 0 & {n_{\bm{q}}^{(\tau) --}}(\bm{x}^{(\tau)})
        \end{array} \Bigg) \Bigg] },
\end{equation}
with
\begin{subequations}
\begin{align}
{n_{\bm{q}}^{(\tau) ++}}(\bm{x}^{(\tau)}) 
&= 
e^{\di \varphi_{l_{\tau}}} {a_{\bm{q} }^{(\tau) ++}}(\bm{x}^{(\tau)}) r(\beta), 
\\
{n_{\bm{q}}^{(\tau) --}}(\bm{x}^{(\tau)}) 
&= 
e^{\di \varphi_{l_{\tau}}} {a_{\bm{q} }^{(\tau) --}}(\bm{x}^{(\tau)}) r^*(\beta). 
\end{align}
\end{subequations}
Since the two $y$-simplex blocks yield identical overlaps, the sign of the total 
overlap is fixed by the sign of any of them.

\paragraph{Rotated Density and Pairing Tensors}
In the $y$-simplex basis, the density matrix acquires a diagonal block structure
\begin{equation}
\rho_{\bm{q}}^{(\tau)}(\bm{x}^{(\tau)}) = 
\Bigg( \begin{array}{cc}
        \rho_{{\bm{q}}}^{(\tau) ++}(\bm{x}^{(\tau)}) & 0 \\
        0 & \rho_{{\bm{q}}}^{(\tau) --} (\bm{x}^{(\tau)})
        \end{array} \Bigg),
\label{eq:density_matrix}
\end{equation}
where
\begin{subequations}
\begin{align}
 \rho_{{\bm{q}}}^{(\tau) ++}(\bm{x}^{(\tau)})&= 
 e^{\di \varphi_{l_{\tau}}}  r(\beta)  
 v_{\bm{q}}^{(\tau)}   \Big[{a_{\bm{q}}^{(\tau) --}}(\bm{x}^{(\tau)})
 \Big]^{-1}  v_{\bm{q}}^{(\tau) \dagger}, \\ 
 \rho_{\bm{q}}^{(\tau) --}(\bm{x}^{(\tau)}) &=
  e^{\di \varphi_{l_{\tau}}}  
 r^*(\beta) v_{\bm{q}}^{(\tau) *} 
 \Big[{a_{\bm{q}}^{(\tau) ++}}(\bm{x}^{(\tau)})
 \Big]^{-1}  v_{\bm{q}}^{(\tau) T}.
\end{align}
\end{subequations}
On the other hand, the pairing tensor acquires an off-diagonal block structure
\begin{equation}
\kappa^{(\tau)}_{\bm{q}}(\bm{x}^{(\tau)}) 
= 
\Bigg( \begin{array}{cc}
0 & \kappa_{\bm{q}}^{(\tau) +-}(\bm{x}^{(\tau)}) \\
    \kappa_{\bm{q}}^{(\tau) -+}(\bm{x}^{(\tau)}) & 0 
\end{array} \Bigg),
\end{equation}
where
\begin{subequations}
\begin{align}
\kappa_{\bm{q}}^{(\tau) +-}(\bm{x}^{(\tau)})
& = -e^{\di \varphi_{l_{\tau}}} r(\beta) v_{\bm{q}}^{(\tau)} 
\Big[{a_{\bm{q}}^{(\tau) --}}(\bm{x}^{(\tau)})\Big]^{-1} u_{\bm{q}}^{(\tau) \dagger}, 
\\ 
\kappa_{\bm{q}}^{(\tau) -+}(\bm{x}^{(\tau)}) 
&= e^{\di \varphi_{l_{\tau}}} r^*(\beta)  v_{\bm{q}}^{(\tau) *} 
\Big[{a_{\bm{q}}^{(\tau) ++}}(\bm{x}^{(\tau)}) \Big]^{-1}  u_{\bm{q}}^{(\tau) T}.
\end{align}
\end{subequations}
Similarly, 
\begin{equation}
\kappa^{*(\tau)}_{\bm{q}}(\bm{x}^{(\tau)}) 
= 
\Bigg( \begin{array}{cc}
0 & \kappa_{\bm{q}}^{* (\tau) +-}(\bm{x}^{(\tau)}) \\
    \kappa_{\bm{q}}^{* (\tau) -+}(\bm{x}^{(\tau)}) & 0 
\end{array} \Bigg),
\end{equation}
with
\begin{subequations}
\begin{align}
\kappa_{\bm{q}}^{* (\tau) +-}(\bm{x}^{(\tau)})
& = -e^{-\di \varphi_{l_{\tau}}} r^*(\beta) u_{\bm{q}}^{(\tau) *} 
\Big[{a_{\bm{q}}^{(\tau) ++}}(\bm{x}^{(\tau)})\Big]^{-1} v_{\bm{q}}^{(\tau) T}, 
\\ 
\kappa_{\bm{q}}^{* (\tau) -+}(\bm{x}^{(\tau)}) 
&= e^{-\di \varphi_{l_{\tau}}} r(\beta)  u_{\bm{q}}^{(\tau)} 
\Big[{a_{\bm{q}}^{(\tau) --}}(\bm{x}^{(\tau)}) \Big]^{-1}  v_{\bm{q}}^{(\tau) \dagger}.
\end{align}
\end{subequations}


\subsubsection{Making Use of the Symmetries}
\label{subsubsec:symmetries}

The expansion in the $y$-simplex basis enables us to reduce the computational 
cost by making all matrices block-diagonal. The computational cost can further 
be reduced by exploiting the symmetries in rotational angle $\beta$ and gauge 
angle $\varphi_{l_{\tau}}$:
\begin{itemize}
\item For reflection-symmetric configurations ($q_{30}=0$), all quantities are 
symmetric around $\beta = \pi/2$. Consequently, the projection interval can be 
reduced to $\beta \! \in \! [0, \pi/2]$. This feature is automatically implemented 
for all reflection-symmetric configurations.
\item The projection interval in gauge angle $\varphi_{l_{\tau}}$ can always be 
reduced to $\varphi_{l_{\tau}} \! \in \! [0, \pi]$ due to the number-parity symmetry 
of an HFB state. In addition, using symmetries of the two simplex blocks, we 
have
\begin{subequations}
\begin{align}
\mathcal{N}^{(\tau)}_{\bm{q}}(\beta,\pi - \varphi_{l_{\tau}}) &=
\Big[\mathcal{N}^{(\tau)}_{\bm{q}}(\beta, \varphi_{l_{\tau}})\Big]^*, \\
\rho_{\bm{q}}^{(\tau) ++}( \beta, \pi-\varphi_{l_{\tau}}) &= 
\Big[\rho_{\bm{q}}^{(\tau) --}(\beta, \varphi_{l_{\tau}})\Big]^*, \\
\rho_{\bm{q}}^{(\tau) --}(\beta,\pi-\varphi_{l_{\tau}}) &= 
\Big[\rho_{\bm{q}}^{(\tau) ++}(\beta,\varphi_{l_{\tau}})\Big]^{*}, \\
\kappa_{\bm{q}}^{(\tau) +-}(\beta,\pi-\varphi_{l_{\tau}}) &= 
-[\kappa_{\bm{q}}^{(\tau) -+}(\beta,\varphi_{l_{\tau}})]^{*}, \\
\kappa^{(\tau) -+}_{\bm{q}}(\beta, \pi-\varphi_{l_{\tau}}) &= 
-\Big[\kappa^{(\tau) +-}_{\bm{q}}(\beta, \varphi_{l_{\tau}})\Big]^{*}, \\
\kappa_{\bm{q}}^{* (\tau) +-}(\beta,\pi-\varphi_{l_{\tau}}) &= 
-[\kappa_{\bm{q}}^{* (\tau) -+}(\beta,\varphi_{l_{\tau}})]^{*}, \\
\kappa^{* (\tau) -+}_{\bm{q}}(\beta, \pi-\varphi_{l_{\tau}}) &= 
-\Big[\kappa^{* (\tau) +-}_{\bm{q}}(\beta, \varphi_{l_{\tau}})\Big]^{*}.
\end{align}
\end{subequations}
Consequently, only quantities within the interval 
$\varphi_{l_{\tau}} \in [0, \pi/2]$ are explicitly calculated.
\end{itemize}

\subsubsection{Densities in the Coordinate-Space Representation}
\label{subsubsec:coordinate_space}

The expressions \eqref{eq:rotated_density}
- \eqref{eq:rotated_pairing_tensor*} 
for the rotated (transition) density and pairing tensors are written in 
the configuration space, that is, the quantities $U_{\gras{q}}^{(\tau)}$, $V_{\gras{q}}^{(\tau)}$, etc., 
are matrices. When using Skyrme EDFs, the coordinate-space 
representation is also especially useful.
 
\paragraph{General Expressions} In the coordinate-space representation,
the full one-body density matrix for particle type $\tau$
can be written as 
\begin{align}
\begin{split}
\rho_{\bm{q}}^{(\tau)}(\bm{r}\sigma,\bm{r'}\sigma') &= 
\frac{1}{2}\rho_{\bm{q}}^{(\tau)}(\bm{r},\bm{r'})\delta_{\sigma \sigma'}
 \\ & + \frac{1}{2} \sum_{\mu}
\braket{\sigma|\hat{\sigma}_{\mu}|\sigma'}s_{\bm{q}, \mu}^{(\tau)}(\bm{r},\bm{r'}),
\end{split}
\end{align}
where $\rho_{\bm{q}}^{(\tau)}(\bm{r},\bm{r'})$ is the non-local one-body particle 
density
\begin{equation}
\rho_{\bm{q}}^{(\tau)}(\bm{r},\bm{r'}) = \sum_{\sigma} 
\rho_{\bm{q}}^{(\tau)}(\bm{r}\sigma,\bm{r'}\sigma)
\end{equation}
and $s_{\bm{q},\mu}^{(\tau)}(\bm{r},\bm{r'})$ is the $\mu$ component of the non-local 
one-body spin density
\begin{equation}
s_{\bm{q}, \mu}^{(\tau)}(\bm{r},\bm{r'}) = \sum_{\sigma \sigma'} 
\rho_{\bm{q}}^{(\tau)}(\bm{r}\sigma,\bm{r'}\sigma') 
\braket{\sigma' | \sigma_{\mu} | \sigma}.
\end{equation}
These non-local densities
can be used to generate an auxiliary set of local
densities that will appear in the expression
for the energy density functional.
In particular, the local particle density 
$\rho_{\bm{q}}^{{(\tau)}}({\bm{r}})$, 
the local spin density $\bm{s}_{\bm{q}}^{(\tau)}(\bm{r})$,
the kinetic energy density $\tau_{\bm{q}}^{(\tau)}(\bm{r})$, 
the spin kinetic energy density $\bm{T}_{\bm{q}}^{(\tau)}(\bm{r})$, 
the current density $\bm{j}_{\bm{q}}^{(\tau)}(\bm{r})$, 
and the spin current density $\tensor{J}_{\bm{q}}^{(\tau)}(\bm{r})$
read
\begin{subequations}
\begin{align}
\rho_{\bm{q}}^{(\tau)}(\bm{r}) &= \rho_{\bm{q}}^{(\tau)}(\bm{r},\bm{r}), 
\label{eq:particle_density} \\
\bm{s}_{\bm{q}}^{(\tau)}(\bm{r}) &= \bm{s}_{\bm{q}}^{(\tau)}(\bm{r},\bm{r}), 
\label{eq:spin_density} \\
\tau_{\bm{q}}^{(\tau)}(\bm{r}) &= \nabla \cdot \nabla'
\rho_{\bm{q}}^{(\tau)}(\bm{r},\bm{r'})\rvert_{\bm{r'}=\bm{r}},
\label{eq:kinetic_density} \\
T_{\bm{q},\mu}^{(\tau)}(\bm{r}) &= \nabla \cdot \nabla'
s_{\bm{q},\mu}^{(\tau)}(\bm{r},\bm{r'})\rvert_{\bm{r'}=\bm{r}}, 
\label{eq:spin_kinetic_density} \\
\bm{j}_{\bm{q}}^{(\tau)}(\bm{r}) &= \frac{1}{2\di}
(\nabla - \nabla') \rho_{\bm{q}}^{(\tau)}(\bm{r},\bm{r'})
\rvert_{\bm{r'}=\bm{r}}, 
\label{eq:current_density} \\
J_{\bm{q}, \mu \nu}^{(\tau)}(\bm{r}) &= \frac{1}{2\di}
(\nabla_{\mu} - \nabla'_{\mu}) 
s_{\bm{q}, \nu}^{(\tau)}(\bm{r},\bm{r'})\rvert_{\bm{r'}=\bm{r}},
\label{eq:spin_current_density}.
\end{align}
\end{subequations}
Furthermore, the non-local pairing densities
for particle type $\tau$ are defined through 
the corresponding pairing tensors as 
\begin{subequations}
\begin{align}
\tilde{\rho}_{\bm{q}}^{(\tau)}(\bm{r}\sigma,\bm{r'}\sigma') 
&= (-2\sigma') \kappa_{\bm{q}}^{(\tau)} (\bm{r}\sigma,\bm{r'}\!-\!\sigma'), \\
\tilde{\rho}_{\bm{q}}^{* (\tau)}(\bm{r}\sigma,\bm{r'}\sigma') 
&= (-2\sigma') \kappa_{\bm{q}}^{* (\tau)} (\bm{r}\sigma,\bm{r'}\!-\!\sigma').
\end{align}
\end{subequations}
They can be equivalently expanded as
\begin{align}
\begin{split}
\tilde{\rho}_{\bm{q}}^{(\tau)}(\bm{r}\sigma,\bm{r'}\sigma') &= 
\frac{1}{2}\tilde{\rho}_{\bm{q}}^{(\tau)}(\bm{r},\bm{r'})
\delta_{\sigma \sigma'} \\ & + \frac{1}{2} \sum_{\mu}
\braket{\sigma|\hat{\sigma}_{\mu}|\sigma'}
\tilde{s}_{\bm{q},\mu}^{(\tau)}(\bm{r},\bm{r'}).
\end{split}
\end{align}
However, only local pairing densities will be 
considered in the pairing term of the energy
density functional
\begin{subequations}
\begin{align}
\tilde{\rho}_{\bm{q}}^{(\tau)}(\bm{r}) &= 
\tilde{\rho}_{\bm{q}}^{(\tau)}(\bm{r},\bm{r}),
\label{eq:pairing_density} \\
\tilde{\rho}_{\bm{q}}^{*(\tau)}(\bm{r}) &= 
\tilde{\rho}_{\bm{q}}^{* (\tau)}(\bm{r},\bm{r}).
\label{eq:pairing_density*} 
\end{align}
\end{subequations}

Formally, equations \eqref{eq:particle_density}~-~\eqref{eq:spin_current_density}
and \eqref{eq:pairing_density}~-~\eqref{eq:pairing_density*} look identical 
regardless of whether $\rho_{\bm{q}}^{(\tau)}(\bm{r}\sigma,\bm{r'}\sigma')$ is the 
diagonal one-body density matrix,
\begin{equation}
\rho_{\bm{q}}^{(\tau)}(\bm{r}\sigma,\bm{r}'\sigma') 
\equiv 
\frac{\braket{\Phi_{\bm{q}}|c^{\dagger}(\bm{r}'\sigma'\tau)c(\bm{r}\sigma\tau)|\Phi_{\bm{q}}}}{\braket{\Phi_{\bm{q}}|\Phi_{\bm{q}}}}
\label{eq:density_SREDF}
\end{equation}
or the rotated (transition) one-body density,
\begin{equation}
\rho_{\bm{q}}^{(\tau)}(\bm{r}\sigma,\bm{r'}\sigma'; \eta) 
\equiv 
\frac{\braket{\Phi_{\bm{q}}|c^{\dagger}(\bm{r}'\sigma'\tau)c(\bm{r}\sigma\tau)\mathcal{R}[\eta]|\Phi_{\bm{q}}}}{\braket{\Phi_{\bm{q}}|\mathcal{R}[\eta]|\Phi_{\bm{q}}}},
\label{eq:density_MREDF}
\end{equation}
where $c^{\dagger}(\bm{r}'\sigma'\tau)$ and $c(\bm{r}\sigma\tau)$
are the creation and the annihilation operator for particle $\tau$
corresponding to the single-particle basis of choice,
$\mathcal{R}$ is the transformation (rotation) operator
related to the symmetry being restored,
and $\eta$ denotes a set of real 
numbers parametrizing the elements of the symmetry group(s) related to
the transformation $\mathcal{R}$
(that is, in our case, $\eta \equiv \gras{x}^{(\tau)}$). 
The main difference is that for diagonal one-body density matrix all local densities are 
real-valued if axial-symmetry is enforced. 
On the other hand, the densities
stemming from the latter 
matrix are generally complex-valued \cite{rohozinski2010selfconsistent}. For 
completeness, we give the explicit expressions for the densities and currents
\eqref{eq:particle_density}~-~\eqref{eq:spin_current_density} and 
\eqref{eq:pairing_density}~-~\eqref{eq:pairing_density*} in 
\ref{sec:app_densities}.

\paragraph{Time-Odd Densities and Symmetry Restoration}
Within the HFB theory, the local densities
$\rho_{\bm{q}}^{(\tau)}$, $\tau_{\bm{q}}^{(\tau)}$, 
and $\tensor{J}_{\bm{q}}^{(\tau)}$ are even, 
while $\bm{s}_{\bm{q}}^{(\tau)}$, $\bm{T}_{\bm{q}}^{(\tau)}$, and 
$\bm{j}_{\bm{q}}^{(\tau)}$ are odd 
under the time-reversal transformation
\cite{engel1975timedependent}. When the HFB state $\ket{\Phi_{\bm{q}}}$ in 
\eqref{eq:density_SREDF} is time-even, as is the case for even-even nuclei
at the SR-EDF level, the  
$\rho_{\bm{q}}^{(\tau)}(\gras{r}\sigma,\gras{r}'\sigma')$
matrix is time-even
as well. Consequently, one can 
show that in such cases $\bm{s}_{\bm{q}}^{(\tau)}(\bm{r}) = \bm{T}_{\bm{q}}^{(\tau)}(\bm{r}) = 
\bm{j}_{\bm{q}}^{(\tau)}(\bm{r}) = 0$
and the corresponding energy contributions
vanish identically. 
Furthermore, blocking calculations for
odd nuclei in {\hfbtho} are implemented
in the equal filling approximation
\cite{perez-martin2008microscopic}, which
enforces the conservation of time-reversal
symmetry. Therefore, the time-odd densities
do not contribute in this case either.

However, the situation is generally different for transition densities of Eq.~\eqref{eq:density_MREDF}, such as 
the gauge- and Euler-rotated densities appearing at the MR-EDF level \cite{rohozinski2010selfconsistent}. Most 
importantly, the transition densities are
generally not Hermitian. Consequently, even if the HFB state is time-even, the time-odd densities and the corresponding
energy contributions may 
not vanish identically. In the particular case of 
particle number projection (PNP), one can 
show that the one-body density matrix is symmetric in the oscillator basis and that, as a result, 
the spin density transforms under 
the time-reversal as $\hat{T}\bm{s}_{\bm{q},\mu}^{(\tau)}(\gras{r},\gras{r}') \! =  \!
-\bm{s}_{\bm{q},\mu}^{(\tau)}(\gras{r},\gras{r}')$. This property ensures that the spin 
density vanishes identically when the reference state is time-even. However, this result is specific to the case of PNP alone. For the angular momentum projection (AMP) 
or the combined PNP and AMP,
all time-odd densities are generally non-zero and contribute to 
the projected energy (or any other observable).

\subsubsection{Rotated Energy Density Functional}
\label{subsubsec:hamiltonian_kernel}
\paragraph{Rotated Hamiltonian Kernel}The rotated Hamiltonian kernel is a functional of the rotated 
density and rotated pairing tensors.
It corresponds to a spatial integral
of the rotated energy density functional
\begin{equation}
\mathcal{H}_{\bm{q}}(\bm{x}) [\rho, \kappa, \kappa^*] = \int d^3\bm{r}\,\mathcal{E}_{\bm{q}}(\bm{r}; \bm{x})[\rho, \kappa, \kappa^*],
\end{equation}
where $\bm{x} \equiv \{ \bm{x^{(\tau=n)}}, \bm{x^{(\tau=p)}} \}$. Version {\codeversion} of \pr{hfbtho} implements the 
restoration of symmetries for Skyrme-based EDFs only.

The total EDF can be decomposed into the 
particle-hole (Skyrme) part and the
particle-particle (pairing) part
\begin{equation}
\mathcal{E}_{\bm{q}}(\bm{r};\bm{x}) = 
\mathcal{E}_{\bm{q}}^{\text{Sky}}(\bm{r};\bm{x})+
\mathcal{E}_{\bm{q}}^{\text{pair}}(\bm{r};\bm{x}),
\label{eq:totedf}
\end{equation}
where
\begin{equation}
\mathcal{E}_{\bm{q}}^{\text{Sky}}(\bm{r};\bm{x})= \mathcal{E}_{\bm{q}}^{\text{kin}}(\bm{r}; \bm{x})  + \mathcal{E}_{\bm{q}}^{\text{Cou}}(\bm{r};\bm{x})  +\mathcal{E}_{\bm{q}}^{\text{pot}}(\bm{r};\bm{x}).
\label{eq:skyedf}
\end{equation}
Note that functional dependencies on the rotated density and pairing tensors were dropped for compactness
on each side of Eqs.~\eqref{eq:totedf}
and \eqref{eq:skyedf}. The kinetic term simply reads
\begin{subequations}
\begin{align}
\mathcal{E}_{\bm{q}}^{\text{kin}}(\bm{r};\bm{x})&= \sum_{\tau = n, p}
\frac{\hbar^2}{2m} \tau_{\bm{q}}^{(\tau)}(\bm{r}; \bm{x}).
\end{align}
\end{subequations}
The Coulomb term can be decomposed into the
direct
and the exchange part,
$\mathcal{E}_{\bm{q}}^{\text{Cou}}(\bm{r};\bm{x}) =
\mathcal{E}_{\bm{q}}^{\text{Cou},\text{dir}}(\bm{r};\bm{x})+
\mathcal{E}_{\bm{q}}^{\text{Cou},\text{exc}}(\bm{r};\bm{x})$.
The direct contribution is calculated as
\begin{equation}
\mathcal{E}_{\bm{q}}^{\text{Cou},\text{dir}}(\bm{r};\bm{x}) =
\frac{1}{2} \int \,d^3\bm{r'}\frac{\rho_{\bm{q}}^{(p)}(\bm{r};\bm{x})\rho_{\bm{q}}^{(p)}(\bm{r'})}{|\bm{r}-\bm{r'}|},
\end{equation}
while the exchange contribution is calculated
in the local Slater approximation
\begin{equation}
\mathcal{E}_{\bm{q}}^{\text{Cou},\text{exc}}(\bm{r};\bm{x}) =
-\frac{3e^2}{4}\left(\frac{3}{\pi} \right)^{1/3}
\Big[\rho_{\bm{q}}^{(p)}(\bm{r};\bm{x})\Big]^{4/3}.
\end{equation}
Note that the pairing contribution of the
Coulomb interaction has been omitted and the Coulomb potential is computed with 
the non-rotated density to save computational time. The resulting error is less 
than 100 keV on the $J=10$ state of Table~\ref{tab:AMP_HFB}.

Furthermore, the Skyrme pseudopotential term
can also be decomposed into two contributions
\begin{equation}
\mathcal{E}_{\bm{q}}^{\text{pot}}(\bm{r};\bm{x}) = \sum_{t=0,1}
\Big[\mathcal{E}_{\bm{q},t}^{\text{pot},\text{even}}(\bm{r};\bm{x})
+\mathcal{E}_{\bm{q}, t}^{\text{pot},\text{odd}}(\bm{r};\bm{x})\Big],
\label{eq:potential_term}
\end{equation}
where the former is built from time-even densities and currents only, while the
latter is built from time-odd 
densities and currents only. Of course, both
contributions are themselves time-even
by construction. 
Furthermore, the summation over $t$ 
in Eq.~(\ref{eq:potential_term})
reflects the 
coupling of neutron and proton
densities and currents 
into the isoscalar ($t = 0$) and the
isovector ($t = 1$) channel, i.e.
\begin{align}
\begin{split}
\rho_{\bm{q}, 0}(\bm{r};\bm{x})  &=
 \rho_{\bm{q}}^{(n)}(\bm{r};\bm{x})  + \rho_{\bm{q}}^{(p)}(\bm{r};\bm{x}),  \\
\rho_{\bm{q}, 1}(\bm{r};\bm{x})  &= 
\rho_{\bm{q}}^{(n)}(\bm{r};\bm{x})  - \rho_{\bm{q}}^{(p)}(\bm{r};\bm{x}),
\end{split}
\end{align}
and equivalently for other
densities and currents.
The time-even contribution to the EDF then reads
\begin{align}
\begin{split}
\mathcal{E}_{\bm{q}, t}^{\text{pot},\text{even}}(\bm{r};\bm{x}) &= 
C_{\bm{q},t}^{\rho \rho}(\bm{r};\bm{x}) \rho_{\bm{q},t}^2(\bm{r};\bm{x}) \\ & +
C_t^{\rho \Delta \rho} \rho_{\bm{q},t}(\bm{r};\bm{x}) \Delta \rho_{\bm{q},t}(\bm{r};\bm{x}) \\& 
+ C_t^{\rho \tau} \rho_{\bm{q},t}(\bm{r};\bm{x}) \tau_{\bm{q},t}(\bm{r};\bm{x}) \\&
+  C_t^{\rho \nabla J} 
\rho_{\bm{q},t}(\bm{r};\bm{x}) \nabla \cdot \tensor{\bm{J}}_{\bm{q},t}(\bm{r};\bm{x}) \\ & + C_t^{JJ} \sum_{\mu \nu} J_{\bm{q},t, \mu \nu}(\bm{r};\bm{x})
J_{\bm{q},t, \mu \nu}(\bm{r};\bm{x}),
\label{eq:skyrme_even}
\end{split}
\end{align}
and the time-odd contribution reads 
\begin{align}
\begin{split}
\mathcal{E}_{\bm{q},t}^{\text{pot},\text{odd}}(\bm{r};\bm{x}) &= C_{\bm{q},t}^{ss}(\bm{r};\bm{x}) \bm{s}_{\bm{q},t}^2(\bm{r};\bm{x}) \\ &
+ C_t^{s \Delta s} \bm{s}_{\bm{q},t}(\bm{r};\bm{x}) \Delta \bm{s}_{\bm{q},t}(\bm{r};\bm{x}) \\
 & + C_t^{sj} \bm{j}^2_{\bm{q},t}(\bm{r};\bm{x})
 \\ & +  C_t^{s \nabla j} \bm{s}_{\bm{q},t}(\bm{r};\bm{x}) \cdot \Big(\nabla \times \bm{j}_{\bm{q},t}(\bm{r};\bm{x})\Big) \\
 & + C_{t}^{sT} \bm{s}_{\bm{q},t}(\bm{r};\bm{x}) \cdot \bm{T}_{\bm{q},t}(\bm{r};\bm{x}).
\label{eq:skyrme_odd}
\end{split}
\end{align}
Note that the coupling constants $C_{\bm{q},t}^{\rho \rho}(\bm{r};\bm{x})$ and
$C_{\bm{q},t}^{ss}(\bm{r};\bm{x})$ are density-dependent.
Furthermore, the last terms in Eqs.~\eqref{eq:skyrme_even}
and \eqref{eq:skyrme_odd} represent
tensor contributions and are set to zero by construction
in a number of Skyrme EDFs. 
The full expressions for coupling constants $C_t$ in terms
of the $(t, x)$ parameters of the Skyrme EDF
are given in \ref{sec:app_couplingconstants}.

Finally, the pairing
term reads
\begin{equation}
\mathcal{E}_{\bm{q}}^{\text{pair}}(\bm{r};\bm{x})
=
\sum_{\tau = n, p}C_{\bm{q}}^{\text{pair}(\tau)}(\bm{r},\bm{x})
\tilde{\rho}_{\bm{q}}^{(\tau)}(\bm{r};\bm{x})\tilde{\rho}_{\bm{q}}^{*(\tau)}(\bm{r};\bm{x}),
\end{equation}
with
\begin{equation}
C_{\bm{q}}^{\text{pair}(\tau)}(\bm{r},\bm{x})
=
\frac{V^{(\tau)}_0}{4}
\left[ 1-V^{(\tau)}_1\left(\frac{\rho_{\bm{q}}(\bm{r};\bm{x})}{\rho_c}\right)\right],
\end{equation}
where $V^{(\tau)}_0$ is the pairing strength
for particle $\tau$, $V^{(\tau)}_1$ controls
the nature of pairing between the pure
volume ($V^{(\tau)}_1=0$) and the pure surface 
($V^{(\tau)}_1=1$) interaction,
and $\rho_c = 0.16$~fm$^{-3}$ is the saturation density
of nuclear matter.

\paragraph{Rotated Hamiltonian Kernel of Density-Dependent
Terms} 
Nearly all parameterizations of Skyrme and 
Gogny EDFs include a density-dependent two-body term.
This term has a strongly repulsive character
and was originally
introduced to reproduce the saturation
property of the nuclear interaction.
However, since it is not linked to a genuine
Hamiltonian operator, its contribution
to the rotated Hamiltonian kernel is ambiguous.
In fact, this contribution can be 
determined only by introducing an additional
prescription \cite{robledo2007particle,robledo2010remarks}.
The choice of prescription will influence the 
calculated projected energies and can therefore be considered
as yet another parameter of a density-dependent EDF. 

A common choice is the 
{\it mixed density} prescription
\begin{equation}
\rho^{(\tau)}_{\bm{q},\text{mix}}(\bm{r};\beta,\varphi_{l_{\tau}})= 
\frac{\braket{\Phi_{\bm{q}} | \hat{\rho}^{(\tau)}(\bm{r}) e^{-\di \beta \hat{J}_y} 
e^{\di \varphi_{l_{\tau}} \hat{\tau}} |
 \Phi_{\bm{q}}}}{\braket{\Phi_{\bm{q}} |\Phi_{\bm{q}}}},
\label{eq:mixed_density}
\end{equation} 
where $\hat{\rho}^{(\tau)}(\bm{r})$ is the one-body density operator for particle 
type $\tau$ at point $\bm{r}$. This 
prescription is motivated by
the expression for the Hamiltonian kernel
of density-independent interactions based on the
generalized Wick theorem.
Moreover, it is the only prescription
on the market satisfying all 
the consistency requirements \cite{robledo2007particle}. Most
importantly, even though the mixed density
\eqref{eq:mixed_density}
is generally complex, 
the resulting projected energies are always real
and invariant under symmetry transformations.
Nevertheless, if a density-dependent term
contains a non-integer power of density, the corresponding energy contribution is generally ill-defined. This issue is essentially insurmountable
and can be circumvented only by using density-dependent
terms with integer powers of density
or a different density prescription. A possible alternative is the {\it projected density} prescription
\begin{equation}
\! \rho^{(\tau)}_{\bm{q},\text{proj}}(\bm{r};\beta)=
\frac{
\braket{\Phi_{\bm{q}} | \hat{\rho}^{(\tau)}(\bm{r}) e^{-\di \beta \hat{J}_y}
\hat{P}^{X} |   
 \Phi_{\bm{q}}}} {\braket{\Phi_{\bm{q}} 
| e^{-\di \beta \hat{J}_y} \hat{P}^{X} |   
 \Phi_{\bm{q}}}},\!
\end{equation} 
which is 
real by construction. Unfortunately, it yields 
non-physical results when used in restoration of spatial symmetries, such 
as the rotational or reflection symmetry \cite{robledo2010remarks}. Nevertheless, a hybrid 
approach is possible in which the mixed density prescription is used when 
restoring spatial symmetries, while the projected density prescription is used 
when restoring the particle number symmetry.
Such an approach has been routinely 
employed in MR-EDF calculations with
Gogny EDFs by the Madrid group \cite{robledo2019mean}.

The Skyrme EDFs included in the current
implementation contain 
two density-dependent terms:
(i) the volume term proportional
to $\rho^{\alpha}(\bm{r})$, where $\alpha$ can
be either integer or non-integer
depending on the EDF, and (ii) the
Coulomb exchange term proportional to $[\rho^{(p)}(\bm{r})]^{4/3}$. In addition,
the pairing interaction is proportional to $\rho(\bm{r})$,
except in the case of the pure volume pairing.
The version \codeversion~of {\hfbtho} implements the mixed density prescription in restoration of 
the rotational, reflection, and particle number symmetry.
However, the
code enables choosing the projected density
prescription in particle number projection
for the volume term with non-integer $\alpha$
and the Coulomb exchange term.


\subsection{HFBTHO Library}
\label{subsec:library}

The code source has been largely refactored to facilitate maintenance and 
future developments. This refactoring included modularizing the code base,
removing obsolescent Fortran statements, and generalizing Fortran 
2003 constructs. In each module, module variables, functions, and subroutines 
are thus explicitly declared as {\tt private} and {\tt public}. Furthermore, arguments 
passed to each function and subroutine have the {\tt intent(in/out/inout)} 
attribute. The internal structure of the code has also been reorganized in 
order to produce an \pr{hfbtho} library. 

Compiling the program generates the following three objects:
\begin{itemize}
\item A Fortran executable called {\tt hfbtho\_main}. The call sequence of 
the program has been modified to provide more flexibility while 
maintaining backward compatibility; refer to Sec.~\ref{subsec:run} for a short
description.
\item A static library {\tt libhfbtho.a}. This library provides, among others, 
the routine {\tt Main\_Program()} with the following call sequence
\begin{verbatim}
Subroutine Main_Program(
   filename_hfbtho,filename_unedf, &
   my_comm_world,my_comm_team, &
   my_n_teams,my_team_color, &
   toggle_output,filename_output, &
   filename_dat,filename_binary)
\end{verbatim}
This routine will execute a full \pr{hfbtho} calculation, possibly across 
different MPI ranks. Its arguments are the following:
\begin{itemize}
\item {\tt filename\_hfbtho}: the name of the input data file containing the 
Namelists. \\
Default: {\tt hfbtho\_NAMELIST.dat};
\item {\tt filename\_unedf}: the name of the input data file containing the 
parameters of the EDF. \\
Default: {\tt hfbtho\_UNEDF.dat};
\item {\tt my\_comm\_world}: the MPI world communicator, typically 
{\tt MPI\_COMM\_WORLD}. When compiling the code without MPI support
({\tt USE\_MPI} = 0), this argument is inactive;
\item {\tt my\_comm\_team}: the MPI communicator used to break the MPI processes
into teams, each of which handles a given \pr{hfbtho} calculation.  Currently, 
distributed parallelism through MPI is only used when restoring broken 
symmetries. Without MPI support, this argument is inactive;
\item {\tt my\_n\_teams}: the number of teams in the calculation. Without MPI support, this argument is inactive;
\item {\tt my\_team\_color}: the team "color"
of the MPI process, i.e., the unique
ID number of the team to which the process has been assigned.
Without MPI support, this argument is inactive;
\item {\tt toggle\_output}: if equal to 0, then no ASCII output is 
recorded on file; if equal to 1, the two files {\tt filename\_output} and {\tt filename\_dat} described below are 
written on disk;
\item {\tt filename\_output}: the name of the ASCII output file 
where the results of the calculation are written. \\
Default: {\tt hfbtho.out};
\item {\tt filename\_dat}: the name of the ASCII output file 
where extended results of the calculations are written. Extended 
results include the self-consistent loop, observables, quasiparticle energies, equivalent 
single-particle energies, and Nilsson labels. \\
Default: {\tt thoout.dat};
\item {\tt filename\_binary}: the name of the binary file where 
the code will store the data needed to restart the iterations. \\
Default: {\tt hfbtho\_output.hel}.
\end{itemize}
\item A Python3 binding. The precise name of the binding will depend on the 
user's system, the Python version, and
the Fortran compiler. 
Assuming the binding is (re)named {\tt hfbtho\_library.so}, it can be used 
directly from a Python environment and provides access to the 
{\tt Main\_Program()} routine. For example:
\begin{verbatim}
from hfbtho_library import Main_Program
\end{verbatim}
or 
\begin{verbatim}
import hfbtho_library
\end{verbatim}

\end{itemize}


\subsection{Other changes}
\label{subsec:other}

\paragraph{SeaLL1 Functional}
The SeaLL1 EDF \cite{bulgac2018minimal} is now available in the code. As a 
reminder, this functional reads
\begin{align}
\begin{split}
\mathcal{E}_{\mathrm{SeaLL1}}(\bm{r}) =&
\frac{\hbar^2}{2m} \Big(\tau^{(n)}(\bm{r}) + \tau^{(p)}(\bm{r})\Big)
\\  + & \sum_{j=0}^2 \Big( a_j \rho_0^{5/3}(\bm{r}) + 
b_j \rho_0^{2}(\bm{r}) + c_j \rho_0^{7/3}(\bm{r}) \Big)~\beta^{2j} \\
+ & \eta_{s} \sum_{\tau=n, p}
\frac{\hbar^{2}}{2m} |\nabla \rho^{(\tau)}(\bm{r})|^2 
+  W_0~\bm{J}_0(\bm{r})\! \cdot\! \nabla\rho_0(\bm{r})  \\
+ & \frac{e^2}{2} \int d^3 \bm{r}' \frac{ \rho^{(p)}(\bm{r}) 
\rho^{(p)}(\bm{r}')}{|\bm{r} - \bm{r}'|}
- \frac{3e^2}{4} \left( \frac{\rho^{(p)}(\bm{r})}{3\pi} \right)^{4/3} \\
+ & \sum_{\tau=n, p} g_{\mathrm{eff}}^{(\tau)}(\bm{r}) |\tilde{\rho}^{(\tau)}(\bm{r})|^2.
\label{eq:seall1}
\end{split}
\hspace{-5mm}
\end{align}
The quantity 
$g_{\mathrm{eff}}^{(\tau)}(\bm{r})$ is the renormalized pairing strength which is 
obtained after regularizing a volume pairing interaction of the form 
$g^{(\tau)}(\bm{r}) = g^{(\tau)}$ \cite{dobaczewski2002contact,bulgac2002local}; see 
\cite{perez2017axially} for details about the implementation of the 
regularization procedure. The SeaLL1 EDF is fully characterized by $11$ 
parameters ($\{ a_j, b_j, c_j \}_{j=0,1,2}, \eta_s, W_0$) in the pairing 
channel and $2$ parameters in the particle-particle channel ($g^{(n)}$ and $g^{(p)}$, with $g^{(n)} = g^{(p)} = g_0$ for 
SeaLL1). Note that, like the UNEDFn functionals, SeaLL1 specifies both the 
particle-hole and the pairing channel.

\begin{figure*}[!ht]
\includegraphics[width=0.98\textwidth]{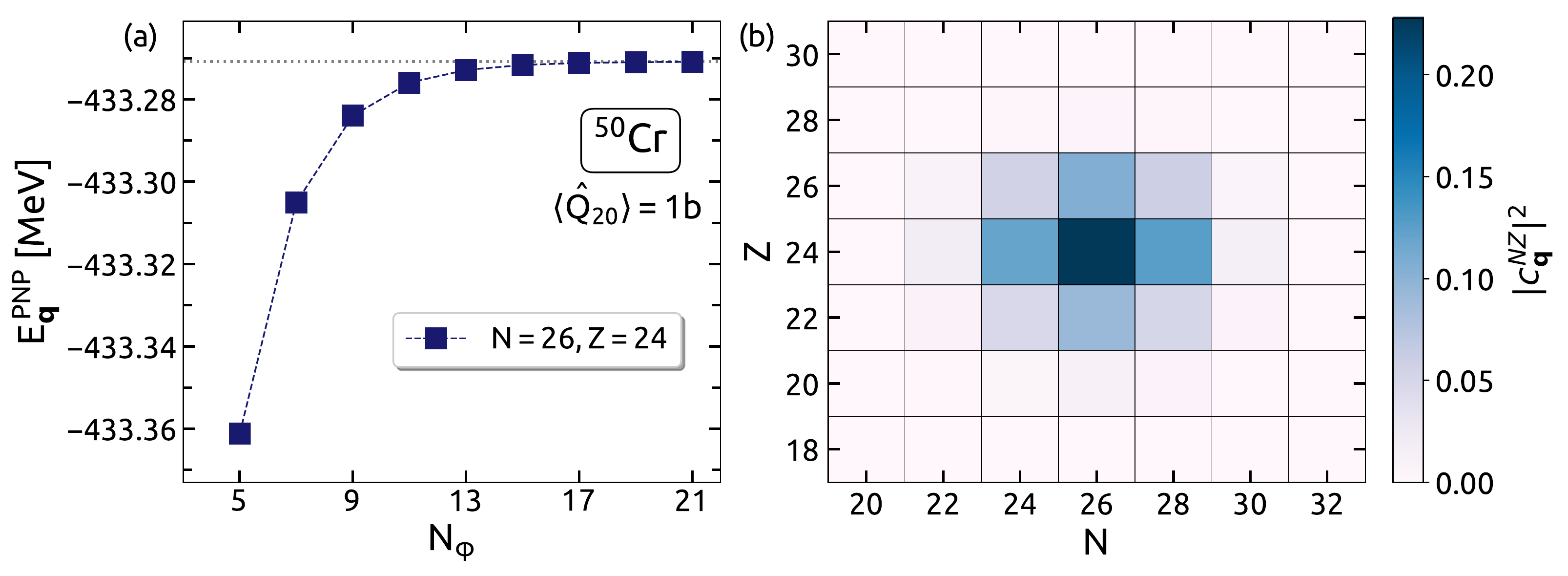}
\caption{Particle number projection in 
the quasiparticle basis for the $\braket{\hat{Q}_{20}} = 1$~b
configuration in $^{50}$Cr. (a): The PNP energy as a function of the number of 
gauge angles $N_{\varphi}$. The dashed horizontal line denotes the fully 
converged solution ($N_{\varphi} = 99$). (b): The decomposition of an HFB state 
onto different numbers of neutrons and protons
for $N_{\varphi} = 15$.}
\label{fig:PNP_convergence}
\end{figure*}

\paragraph{Exact Coulomb}
In previous versions of \pr{hfbtho}, the direct (Hartree) term of the 
Coulomb potential is calculated using the substitution method 
\cite{girod1983triaxial}, the exchange (Fock) term is calculated at the 
Slater approximation, while the pairing term is neglected. As discussed extensively in \cite{stoitsov2013axially}, 
the substitution method can be numerically unstable because of aliasing errors. 
In the current version, we have leveraged the capability to compute mean-field 
and pairing energies from finite-range two-body Gaussian potentials introduced in version 
3.00 to implement an "exact" calculation of the direct, exchange, and pairing  
term of the Coulomb potential. In particular, we follow the technique 
implemented in \cite{dobaczewski2009solution} and discussed in 
\cite{dobaczewski1996meanfield} and by exploiting the identity 
\begin{align}
\begin{split}
\frac{1}{r} 
& = \frac{2}{\sqrt{\pi}} \int_{0}^{+\infty} d\alpha\, e^{-\alpha^2 r^2} \\
& = \frac{2}{L\sqrt{\pi}} \int_{0}^{1} d\xi\, (1-\xi^2)^{-3/2} \exp\left( -\frac{\xi^2 r^2}{L^2(1-\xi^2)} \right),
\end{split}
\end{align}
where we used the change of variable $\alpha = \frac{\xi}{L}(1-\xi^2)^{-1/2}$ 
and $L$ stands for the larger of the two oscillator lengths, 
$L = \max (b_z, b_{\perp} )$. The second integral can be efficiently computed 
with Gauss-Legendre quadrature. If $\omega_i$ and $\xi_i$ are the weights and 
the nodes of Gauss-Legendre quadrature, then we can write
\begin{equation}
\frac{1}{r} = \sum_{i=1}^{N_c} A_i e^{-a_i r^2},
\label{eq:coulomb}
\end{equation}
with $A_i = \frac{2\omega_i}{L\sqrt{\pi}}(1-\xi_{i}^{2})^{-3/2}$ and 
$a_i = \frac{\xi_{i}^{2}}{L^2(1-\xi_{i}^{2})}$.

\paragraph{Overwrite Mode}
The new version of the code provides an option to use the information 
contained in the binary {\tt hfbtho\_output.hel} file to overwrite some of the 
user-defined inputs. This option is activated by setting the energy functional 
to {\tt READ} (instead of the usual {\tt SLY4}, {\tt SKM*}, etc.). In this 
case, the code will overwrite (i) all the parameters of the EDF, 
(ii) the pairing cut-off, (iii) the activation/deactivation of non-standard 
terms such as the center-of-mass correction, tensor terms, or pairing regularization, 
(iv) the parameters of the oscillator basis such as the maximal number of shells 
and oscillator lengths. The code will then redefine the full HO basis to be 
consistent with the one on file.

\paragraph{Bugfix of Blocking Calculations}
In all versions of {\hfbtho} since 2.00d \cite{stoitsov2013axially}, 
there is a bug in the calculations of 
blocked states when the "automatic" mode 
is activated. In this mode, the code 
determines and computes all possible blocking configurations within a $2$ MeV energy 
window around the Fermi level; see Section 4.2 of \cite{stoitsov2013axially} for 
details. In practice, the code loops over all $N$ candidate configurations. 
Occasionally, one of these configurations may diverge, e.g., the particle 
number condition cannot be enforced. When this happened to a configuration 
$1\leq k < N$, the code would simply exit
the loop without trying to compute the remaining 
configurations $k <k'\leq N$. Consequently, the results of the converged calculations were 
correct but some potentially valid configurations were not computed. In 
calculations near the ground state of stable nuclei, this situation occurs very 
rarely; in calculations of very neutron-rich or very deformed nuclei, it may 
happen more frequently. This bug is fixed
in the current version of the code.


\section{Benchmarks and Accuracy}
\label{sec:benchmarks}


\subsection{Particle Number Projection}
\label{subsec:bench_pnp}

As the first illustrative example, we perform
the particle number projection for
a range of quadrupole-deformed configurations in $^{50}$Cr.
Well-converged solutions 
are obtained by 
expanding the HFB states in a spherical HO basis with $N_0 = 8$
shells
and the oscillator length
$b_0 = 1.7621858$ fm. The SIII parametrization of the Skyrme EDF \cite{beiner1975nuclear} is used, alongside
a 
volume ($V_1^{(\tau)} = 0.0$) contact pairing interaction \cite{dobaczewski2002contact} with a $60$ MeV quasiparticle cutoff and pairing 
strengths $V_0^{(n)}\!=\! V_0^{(p)} \!=\!  -190.0$~MeV.
In addition, we employ the mixed density prescription.

\subsubsection{Convergence and Particle Number Decomposition}
We start by testing the convergence of PNP energies
[$E_{\mathbf{q}}^{\text{PNP}} \! \equiv \!E_{\mathbf{q}}^{NZ}$, 
Eq.~\eqref{eq:projected_energy}]
and decomposing an HFB state onto different
numbers of neutrons and protons [$|c_{\mathbf{q}}^{NZ}|^2$,~Eq.~\eqref{eq:decomposition_NZ}].
The quadrupole moment of the reference HFB state
is constrained to $\braket{\hat{Q}_{20}} \! = \! 1$ b, the dipole and the octupole
moment are constrained to zero, while higher
multipole moments are determined self-consistently.
Figure~\ref{fig:PNP_convergence}(a) shows the 
corresponding PNP energy as a function of the number of gauge angles 
$N_{\varphi}$.
An excellent agreement
with the fully converged solution 
(represented by the dashed horizontal line and computed for
$N_{\varphi}=99$) is 
obtained for $N_{\varphi} = 15$. 
The convergence pattern will generally vary for
different HFB states, but at most $N_{\varphi} \! = \! 15$ gauge angles 
should be sufficient for most practical purposes.

Furthermore, Fig.~\ref{fig:PNP_convergence}(b)
shows the decomposition of the same HFB state onto
different numbers of neutrons and protons.
A pronounced maximum is found at the correct 
number of particles, $|c^{N=26, Z=24}_{\bm{q}}|^2 = 0.2278$. Around this point, the distribution
drops sharply in all directions.
For example, the configuration with two protons less
has about twice smaller coefficient,
$|c^{N=26, Z=22}_{\bm{q}}|^2 = 0.1197$, while the
configuration with four protons less has
only $|c^{N=26, Z=20}_{\bm{q}}|^2 = 0.0201$. 
Note that, for this particular configuration, the pairing gaps are $\Delta_n = 1.0901$ MeV and $\Delta_p = 1.1773$ MeV for neutrons
and protons, respectively.

\subsubsection{PNP in Canonical and Quasiparticle Bases}
\label{subsubsec:PNP_can_qp}

The particle number projection in the canonical
basis had been
incorporated to the {\hfbtho} program since its
initial release. On the
other hand, the new version of the program contains the
particle number projection performed in the
quasiparticle basis. The two PNP methods are distinct and can under certain circumstances yield different results. Most notably, a difference will arise if the
underlying HFB calculations enforce a 
cutoff in the quasiparticle space. The introduction
of such a cutoff is a common way to render
the energies convergent for zero-range pairing
interactions 
and is therefore an integral part of Skyrme-EDF calculations
with {\hfbtho} \cite{stoitsov2005axially}.

\begin{figure}[t!ht]
\includegraphics[width=0.48
\textwidth]{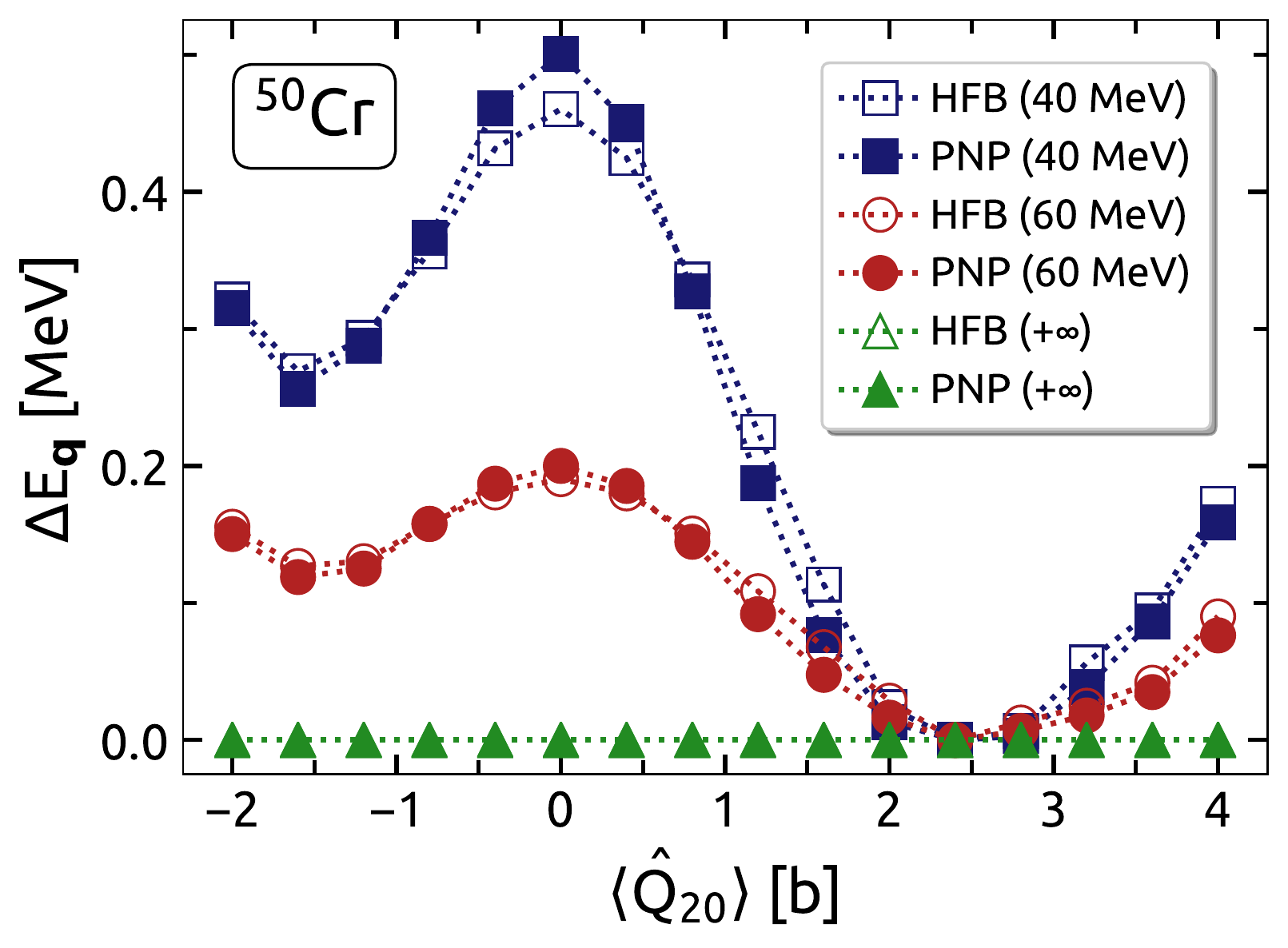}
\caption{The difference between the PNP energies obtained
in the quasiparticle and in the canonical basis,
$\Delta E_{\mathbf{q}}^{\text{PNP}} = 
E_{\mathbf{q}, \text{qps}}^{\text{PNP}}- E_{\mathbf{q}, \text{can}}^{\text{PNP}
}$, for three
different values
of a quasiparticle cutoff: $40$ MeV, 
$60$ MeV, and $6000$ MeV (an infinite cutoff).
The difference in the corresponding HFB energies,
$\Delta E_{\mathbf{q}}^{\text{HFB}} = 
E_{\mathbf{q}, \text{qps}}^{\text{HFB}}
- E_{\mathbf{q}, \text{can}}^{\text{HFB}}$,
is also shown.}
\label{fig:PNP_cutoff}
\end{figure}

To compare the two methods, Fig.~\ref{fig:PNP_cutoff}
shows the difference between the PNP energies obtained 
in the quasiparticle and in the canonical basis,
$\Delta E_{\mathbf{q}}^{\text{PNP}} = 
E_{\mathbf{q}, \text{qps}}^{\text{PNP}}- E_{\mathbf{q}, \text{can}}^{\text{PNP}
}$, for 
three different values of a quasiparticle cutoff.
We consider a range of quadrupole deformations in
$^{50}$Cr, $\braket{\hat{Q}_{20}}  \in [-2.0~\mathrm{b}, 4.0~\mathrm{b}]$, and 
keep the other parameters fixed.
For a relatively low cutoff ($E_\text{cut} = 40$ MeV), the difference
is $\Delta E_{\mathbf{q}}^{\text{PNP}} \le 0.5$ MeV.
For a cutoff value typically used in realistic calculations 
($E_\text{cut} = 60$ MeV), the difference
reduces to $\Delta E_{\mathbf{q}}^{\text{PNP}} \le 0.2$ MeV. Finally,
in the limit of an infinite cutoff ($E_\text{cut} = 6000$ MeV)
the difference between the two methods vanishes. 

In addition, Fig.~\ref{fig:PNP_cutoff} shows the difference
between the HFB energies obtained in the quasiparticle and
in the canonical basis, $\Delta E_{\mathbf{q}}^{\text{HFB}} = 
E_{\mathbf{q}, \text{qps}}^{\text{HFB}}
- E_{\mathbf{q}, \text{can}}^{\text{HFB}}$,
for the three cutoff values. The HFB curves largely follow the corresponding PNP curves, 
corroborating the fact that the discrepancy in projected
energies stems from the initial difference in
HFB states. Finally, an instructive limit
to consider is the case of a
collapsing pairing interaction, which is a common
feature of PNP models that perform variation before
projection \cite{sheikh2021symmetry}. Note 
that the collapse of pairing happens
around $\braket{\hat{Q}_{20}} = 2.5$~b in our calculation. Regardless
of the cutoff, the two PNP methods then
yield the same energy that
also coincides with the HFB energy.

\subsubsection{The Choice of Density Prescription}
\label{subsubsec:prescriptions}

\begin{figure}[!ht]
\includegraphics[width=0.48
\textwidth]{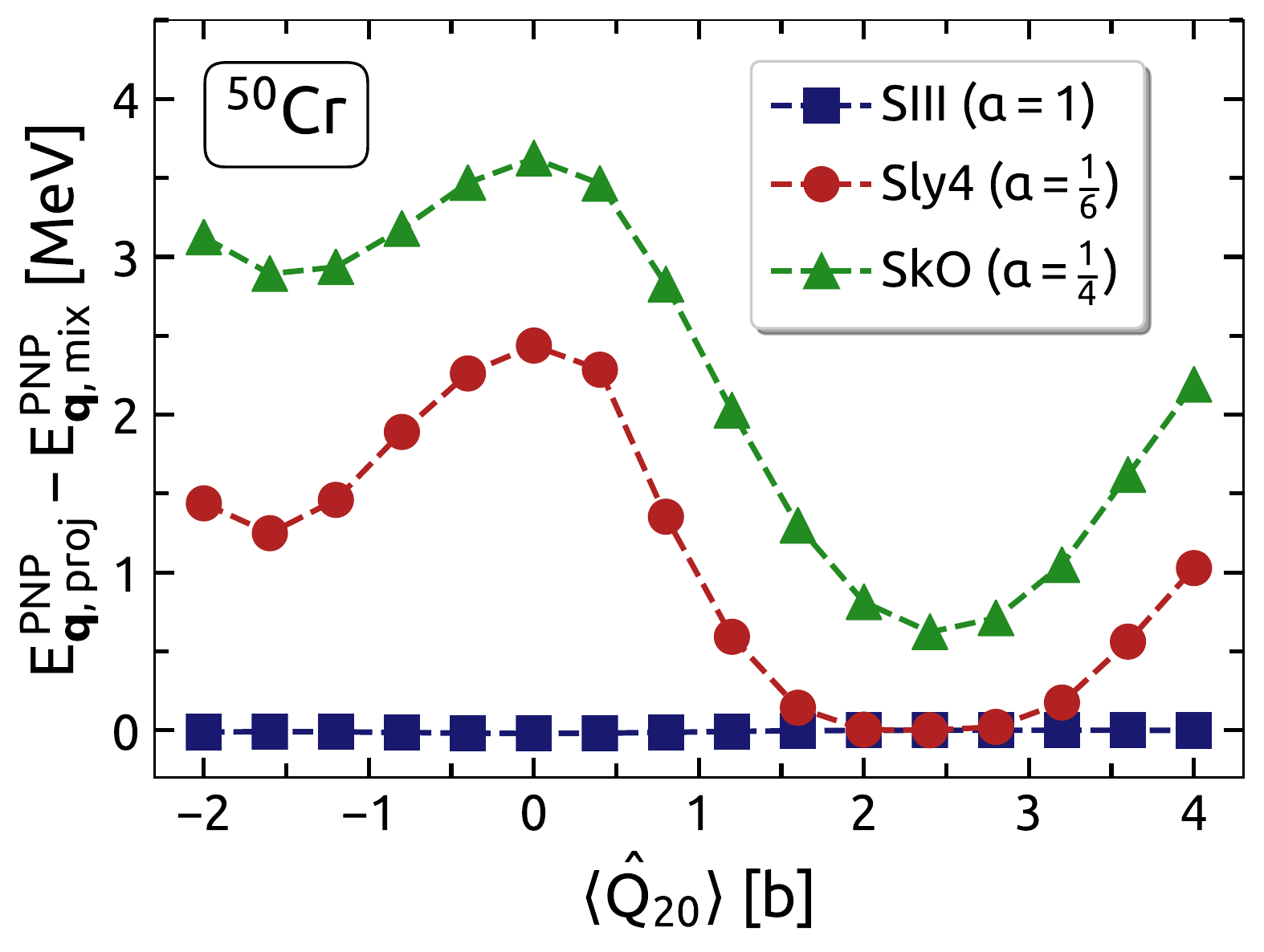}
\caption{The difference between the PNP energies
obtained with the mixed and the projected
density prescription. We consider three Skyrme
EDFs whose volume terms depend on different
powers of density $\alpha$.}
\label{fig:PNP_prescriptions}
\end{figure}

As discussed in Sec.~\ref{subsubsec:hamiltonian_kernel},
the new implementation of PNP enables the choice of
density prescription for the parts of
an EDF that depend on non-integer powers of density.
In order to quantify the consequences of this choice,
Fig.~\ref{fig:PNP_prescriptions} shows the difference 
between the PNP energies obtained with the mixed and 
the projected density prescription. We consider
three Skyrme EDFs whose volume terms
depend on different powers of density $\alpha$:
SIII ($ \alpha = 1$) \cite{beiner1975nuclear}, 
Sly4 ($ \alpha = \frac{1}{6}$) \cite{chabanat1998skyrme}, and SkO ($ \alpha = \frac{1}{4}$) \cite{reinhard1999shape}. 
For all three EDFs, the Coulomb exchange term depends
on the $4/3$-th power of the proton density.

For SIII, the entire difference between the two prescriptions
lies in the Coulomb exchange term. In $^{50}$Cr, this difference amounts to
about $0.1 \%$ of the term, or about $0.01$ MeV, and is
therefore not visible in Fig.~\ref{fig:PNP_prescriptions}. On the
other hand, for Sly4 and SkO an
additional difference in
the volume term comes into play.
The difference in this term amounts to about $0.1 \%$ as well, but it translates
to a sizeable absolute difference of $2 - 3$ MeV. 
Again, the two prescriptions yield the same 
result in the limit of a collapsing
pairing interaction (around $\braket{\hat{Q}_{20}} = 2.5$~b). 
We note that the difference
from density prescriptions does {\it not} 
scale with nuclear mass and
that it remains of comparable magnitude
even in the heaviest nuclei.  

Unfortunately, to the best of our knowledge,
there are no published comparisons of 
PNP energies obtained with different density
prescriptions. However, Ref.~\cite{valor1997new} contains the 
comparison between the PNP dynamic moments of
inertia obtained with the mixed and the projected
density prescription, using a Gogny EDF and
the Lipkin-Nogami approximation. 
The reported difference is sizeable and generally of the 
order of a few percent.

\subsubsection{Benchmarking Against HFODD}
\label{subsubsec:PNP_HFODD}

To further verify our implementation, we tested the PNP 
results of {\hfbtho} against results obtained with {\hfodd}. Since the latest release of the code 
\cite{dobaczewski2021solution} cannot project on both protons and neutrons and does not give a 
full breakdown of the projected 
energy, we use for our benchmark a recent, still unpublished, 
modification of the {\hfodd} solver based on version 2.73 
\cite{schunck2017solution}. In this version, PNP is implemented in the canonical 
basis and the results must thus be tested against the original {\hfbtho} 
implementation \cite{stoitsov2005axially}. As demonstrated in Section 
\ref{subsubsec:PNP_can_qp}, this implementation of PNP (in the canonical basis) 
gives the same results as the new implementation (in the quasiparticle
basis) for infinite cutoffs. 

\begin{table}[!ht]
\center
\caption{\label{tab:PNP} The breakdown of the PNP energy (in MeV) 
of the $\braket{\hat{Q}_{20}} = 1$~b configuration
in $^{50}$Cr, obtained with the {\hfbtho} and {\hfodd} solvers.
A spherical HO basis with $N_0 = 12$ shells and the 
SIII EDF were used; see text for more details on the parameters of the
calculation.}
{\renewcommand{\arraystretch}{1.15} 
\begin{tabular}{ l l l l }
   &  \phantom{-111} {\hfbtho} &  \phantom{-111} {\hfodd} \\
\hline \noalign{\smallskip}
$E_{\rm kin}^{(n)}$  & \phantom{-1}466.23612{\htho 4} & \phantom{-1}466.23612{\hodd 3} \\
$E_{\rm kin}^{(p)}$  & \phantom{-1}415.93724{\htho 4} & \phantom{-1}415.93724{\hodd 3} \\
\hdashline
$E^{\rho\rho}$       &           -1701.7762{\htho 20} &           -1701.7762{\hodd 17} \\
$E^{\rho\tau}$       & \phantom{-1}201.41093{\htho 5} & \phantom{-1}201.41093{\hodd 4} \\
$E^{\rho\Delta\rho}$ & \phantom{-1}126.14195{\htho 9} & \phantom{-1}126.14195{\hodd 8} \\
$E^{\rho\nabla J}$   & \phantom{11}-39.203075         & \phantom{11}-39.203075         \\
\hdashline
$E_{\rm pair}^{(n)}$ & \phantom{111}-0.333798         & \phantom{111}-0.333798         \\
$E_{\rm pair}^{(p)}$ & \phantom{111}-0.981203         & \phantom{111}-0.981203         \\
\hline
$E_{\rm PNP}$        & \phantom{1}-532.568034         & \phantom{1}-532.568034 \\
\end{tabular}
}
\end{table}

Table \ref{tab:PNP} contains a breakdown of the 
PNP energy of the $\braket{\hat{Q}_{20}} \! = \!  1$~b
configuration in $^{50}$Cr, obtained with the {\hfbtho} and {\hfodd} solvers. 
The calculation parameters are the same as those described at the 
beginning of this section, except that
(i) $N_0 = 12$ HO shells are used, (ii)
a surface-volume pairing interaction is used,
and (iii) the Coulomb interaction is entirely neglected. In both {\hfbtho} and 
{\hfodd} calculations, $N_{\varphi} = 15$ gauge angles were used for both neutrons 
and protons. 
The {\hfodd} results correspond to a Gauss quadrature 
characterized by $\texttt{NXHERM}=\texttt{NYHERM}=\texttt{NZHERM}=30$ points. 
The largest difference, for the density-dependent volume term, does not 
exceed $3$ eV.


\subsection{Angular Momentum Projection}
\label{subsec:bench_amp}

Next, we perform the illustrative angular
momentum projection calculations, using 
the same parameters as described at the beginning
of Section \ref{subsec:bench_pnp}.

\begin{figure*}[b!ht]
\includegraphics[width=0.98\textwidth]{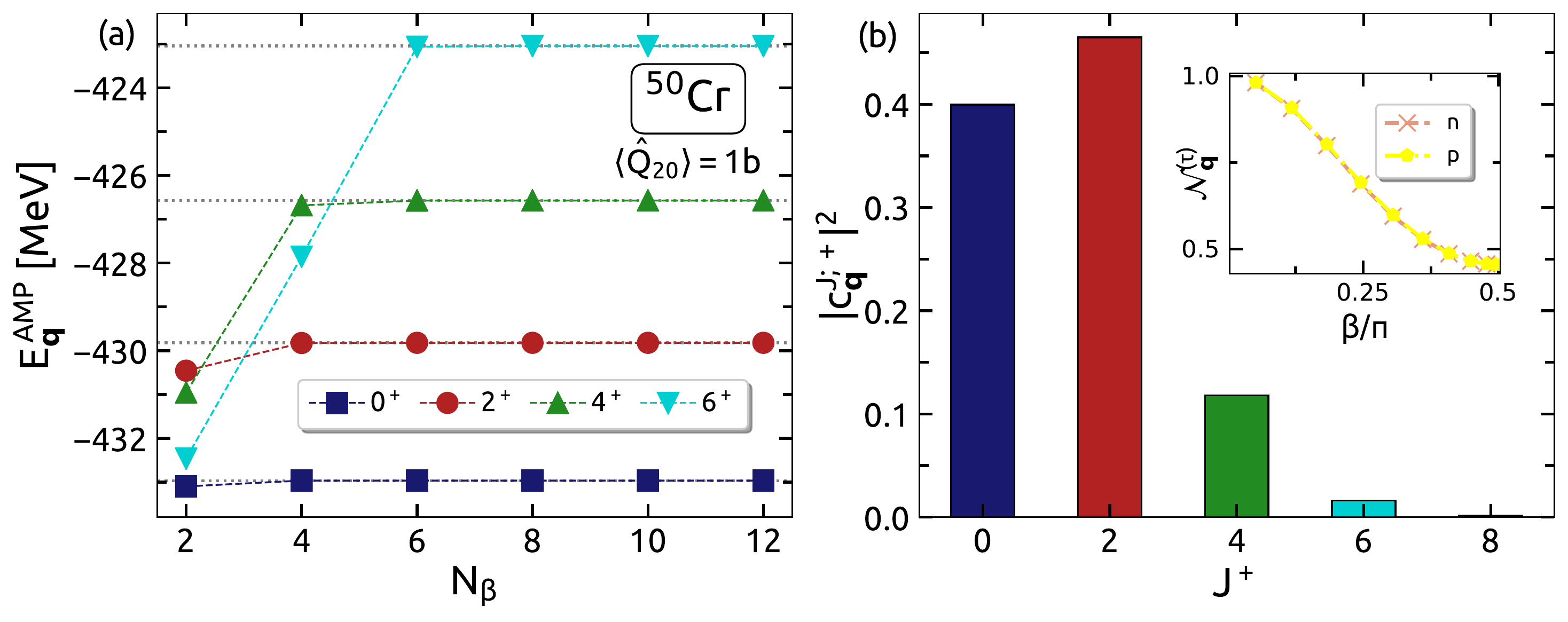}
\caption{Angular momentum projection in 
the spherical HO basis for the $\braket{\hat{Q}_{20}} = 1$~b
configuration in $^{50}$Cr. (a): The AMP energy
of the $J^p = 0^+, 2^+, 4^+$, and $6^+$ state
as a function of the number of 
rotational angles $N_{\beta}$. The dashed horizontal line denotes the fully 
converged solution ($N_{\beta} = 100$). (b): The decomposition of an HFB state 
onto different angular momenta
for $N_{\beta} = 10$. 
The inset shows the corresponding
overlaps for neutrons and protons.}
\label{fig:AMP_convergence}
\end{figure*}

\subsubsection{Convergence of Angular Momentum Decomposition}

To start with, we test the convergence of
AMP energies 
[$E_{\mathbf{q}}^{\text{AMP}} \! \equiv \!E_{\mathbf{q}}^{J; p}$, 
Eq.~\eqref{eq:projected_energy}] and decompose an HFB state
onto different values of angular momenta
[$|c_{\mathbf{q}}^{J; p}|^2$,~Eq.~\eqref{eq:decomposition_J}].
As before, the quadrupole moment of the reference HFB state
is constrained to $\braket{\hat{Q}_{20}} \! = \! 1$ b, the dipole and the octupole
moment are constrained to zero, while higher
multipole moments are determined self-consistently. Fig.~\ref{fig:AMP_convergence}(a)
shows the AMP energies for $J^p = 0^+, 2^+, 4^+$,
and $6^+$ as a function of the number
of rotational angles $N_{\beta}$.
Note that the considered configuration
is reflection-symmetric and thus
only positive-parity states can be obtained.
In turn, the projection interval is reduced
to $\beta \! \in \! [0, \pi/2]$. As expected,
the convergence is faster for lower values
of $J$. For all $J$, an excellent agreement with the 
fully converged solution (represented by the
dashed horizontal lines and computed 
for $N_{\beta} \! = \! 100$) is obtained already
for $N_{\beta} \! = \! 10$. The convergence pattern
will generally depend on the properties
of the HFB state (e.g., the magnitude of
the quadrupole deformation
or whether the parity is broken), as well
as on the value of $J$.
Consequently, in practical applications,
one should verify the convergence
of AMP with respect to $N_{\beta}$.

Furthermore, Fig.~\ref{fig:AMP_convergence}(b) 
shows the decomposition of the same HFB state
onto different values of angular momentum.
The maximum is found for $J=2$, 
$|c_{\mathbf{q}}^{J;+}|^2 = 0.4649$, while
the coefficients for $J \ge 8$ components are negligible.
The inset shows the corresponding
overlaps for both neutrons
and protons
[$\mathcal{N}_{\bm{q}}^{(\tau)}(\beta, 0)$, Eq.~\eqref{eq:overlap}]. The overlaps
for the two types of
particles are very similar: they are real and monotonously 
decrease from $\mathcal{N}_{\bm{q}}^{(\tau)}(0, 0) = 1$ to their respective minimal values
at $\beta = \pi/2$. Since the quadrupole
deformation is rather moderate, the 
overlaps at $\beta = \pi/2$ are still
sizeable. 
Note that the overlaps for $\beta \! \in [\pi/2, \pi]$ can be obtained
by a reflection around the $\beta = \pi/2$
vertical axis; see Section \ref{subsubsec:symmetries}.


\subsubsection{Benchmarking Against HFODD}

In full analogy with the case of PNP discussed in Section 
\ref{subsubsec:PNP_HFODD}, we can benchmark the AMP results obtained with 
{\hfbtho} against the  results obtained with {\hfodd}. The main restriction in 
this case is that {\hfodd} requires the usage of a spherical HO basis. Once 
again, we consider the $\braket{\hat{Q}_{20}} \! = \!  1$~b configuration in 
$^{50}$Cr. The calculation parameters are the same as those described at the 
beginning of Section  \ref{subsec:bench_pnp}, except that (i) the Coulomb
interaction is entirely neglected, (ii) all the higher multipole moments up to 
the eighth order are constrained to zero, and (iii) in order to additionally 
probe the contribution from the tensor term of the functional, we used the SLy5 
parametrization of the Skyrme EDF \cite{chabanat1998skyrme}. In this case, the 
parameterizations of the pairing interaction yields pairing gaps that are much 
smaller than the experimental ones. However, since our goal is simply to 
compare the two codes against one another, this discrepancy is irrelevant. All 
the AMP calculations were performed with $N_{\beta} = 30$ rotational angles 
$\beta \! \in \! [0, \pi]$. 

We compared our results to those generated
with the latest release of {\hfodd}, where the AMP is 
implemented in the Hartree-Fock basis \cite{dobaczewski2021solution}. Because 
the two codes employ different bases, the obtained HFB energies slightly differ and agree within 
$2.2$ keV. For the projected energies, the difference does not exceed
$12$ keV for the range of angular momentum $J\in [0,10]$. Although this test is already very 
encouraging, we can go one step further and test separately each contribution to the 
projected energy. To this end, we use the
same unpublished version of {\hfodd} built on top of the version 2.73 that was employed 
for the PNP benchmark. In that version of the code, the AMP is implemented in 
the HO basis so a closer comparison is 
possible. As expected, we find that the HFB energies 
agree within $1$ eV: $E_{\rm HFB} = -531.370615$ MeV.

\begin{table}[!tb]
\center
\caption{\label{tab:AMP_HFB}  The breakdown of the AMP 
energy (in MeV) of the $\braket{\hat{Q}_{20}} = 1$ b configuration in $^{50}$Cr,
obtained with the {\hfbtho} and {\hfodd} solvers.
Energies for $J=0$ (top) and $J=8$ (bottom)
are shown. A spherical HO basis with $N_0 = 8$ shells and the 
Sly5 EDF were used; see text for more details on the parameters of the
calculation.}
{\renewcommand{\arraystretch}{1.15} 
\begin{tabular}{ l l l l }
$J=0$   & \phantom{-111}{\hfbtho} & \phantom{-111}{\hfodd}  \\
\hline \noalign{\smallskip}
$E_{\rm kin}^{(n)}$  & \phantom{-1}475.8119{\htho 44} & \phantom{-1}475.8119{\hodd 32} \\
$E_{\rm kin}^{(p)}$  & \phantom{-1}418.693{\htho 797} & \phantom{-1}418.693{\hodd 807} \\
\hdashline
$E^{\rho\rho}$       &           -1797.938577         &           -1797.938577         \\
$E^{\rho\tau}$       & \phantom{-1}269.775424         & \phantom{-1}269.775424         \\
$E^{\rho\Delta\rho}$ & \phantom{-1}149.16685{\htho 9} & \phantom{-1}149.16685{\hodd 8} \\
$E^{\rho\nabla J}$   & \phantom{11}-42.0393{\htho 41} & \phantom{11}-42.0393{\hodd 39} \\
$E^{JJ}$             & \phantom{-111}1.213084         & \phantom{-111}1.213084         \\
\hdashline
$E^{ss}$             & \phantom{-111}0.2514{\htho 40} & \phantom{-111}0.2514{\hodd 39} \\
$E^{sj}$             & \phantom{-111}0.28758{\htho 6} & \phantom{-111}0.28758{\hodd 5} \\
$E^{s\Delta s}$      & \phantom{-111}0.11128{\htho 1} & \phantom{-111}0.11128{\hodd 0} \\
$E^{s\nabla J}$      & \phantom{-111}0.13786{\htho 6} & \phantom{-111}0.13786{\hodd 5} \\
$E^{s T}$            & \phantom{-111}0.009186         & \phantom{-111}0.009186         \\
\hdashline
$E_{\rm pair}^{(n)}$     & \phantom{111}-2.84813{\htho 8} & \phantom{111}-2.84813{\hodd 7} \\
$E_{\rm pair}^{(p)}$     & \phantom{111}-4.50788{\htho 7} & \phantom{111}-4.50788{\hodd 5} \\
\hline
$E_{\rm AMP}$        & \phantom{1}-532.30795{\htho 2} & \phantom{1}-532.30795{\hodd 0} \\
\end{tabular}
}
\medskip\\
{\renewcommand{\arraystretch}{1.15} 
\begin{tabular}{ l l l l }
$J=8$   & \phantom{-111}{\hfbtho} & \phantom{-111}{\hfodd}  \\
\hline \noalign{\smallskip}
$E_{\rm kin}^{(n)}$  & \phantom{-1}467.3845{\htho 64} & \phantom{-1}467.3845{\hodd 72} \\
$E_{\rm kin}^{(p)}$  & \phantom{-1}437.860{\htho 544} & \phantom{-1}437.860{\hodd 226} \\
\hdashline
$E^{\rho\rho}$       &           -1812.48{\htho 3313} &           -1812.48{\hodd 2960} \\
$E^{\rho\tau}$       & \phantom{-1}275.246{\htho 980} & \phantom{-1}275.246{\hodd 855} \\
$E^{\rho\Delta\rho}$ & \phantom{-1}148.7249{\htho 58} & \phantom{-1}148.7249{\hodd 62} \\
$E^{\rho\nabla J}$   & \phantom{11}-40.088{\htho 099} & \phantom{11}-40.088{\hodd 112} \\
$E^{JJ}$             & \phantom{-111}0.99776{\htho 0} & \phantom{-111}0.99776{\hodd 3} \\
\hdashline
$E^{ss}$             & \phantom{111}-1.279{\htho 415} & \phantom{111}-1.279{\hodd 386} \\
$E^{sj}$             & \phantom{111}-1.7630{\htho 59} & \phantom{111}-1.7630{\hodd 17} \\
$E^{s\Delta s}$      & \phantom{111}-0.5594{\htho 18} & \phantom{111}-0.5594{\hodd 06} \\
$E^{s\nabla J}$      & \phantom{111}-0.4498{\htho 41} & \phantom{111}-0.4498{\hodd 32} \\
$E^{s T}$            & \phantom{111}-0.07060{\htho 1} & \phantom{111}-0.07060{\hodd 0} \\
\hdashline
$E_{\rm pair}^{(n)}$     & \phantom{111}-1.1595{\htho 25} & \phantom{111}-1.1595{\hodd 44} \\
$E_{\rm pair}^{(p)}$     & \phantom{111}-2.5637{\htho 45} & \phantom{111}-2.5637{\hodd 72} \\
\hline
$E_{\rm AMP}$        & \phantom{1}-527.9638{\htho 05} & \phantom{1}-527.9638{\hodd 95} \\
\end{tabular}
}
\end{table}

Table \ref{tab:AMP_HFB} contains the breakdown of the AMP energy
for angular momentum $J=0$ and $J=8$; see Eqs.~\eqref{eq:skyrme_even}~-~\eqref{eq:skyrme_odd} 
for the definition of each term. For the $J=0$ state, the differences between 
the two codes do not exceed $10$ eV, with most terms agreeing within $2$ eV. Not 
surprisingly, the differences increase a little for the $J=8$ case. However, they are 
still of the order of a few dozens or hundreds of eV, and overall less than $1$ keV. 
Considering the remaining
differences between the two codes -- {\hfodd} works with the 
Cartesian basis and implements the full 3D rotation of wave functions while 
{\hfbtho} works with the cylindrical basis and implements only the rotation in 
the Euler angle $\beta$ -- this benchmark is quite conclusive. 


\subsubsection{AMP in a Deformed Basis}

One of the main advantages of the present implementation of AMP is that 
it can be performed in bases that are not closed under rotation. 
Such deformed (or stretched) bases are often used
in calculations of
potential energy surfaces because they provide a computationally efficient way 
to obtain precise representations of arbitrarily deformed HFB configurations. The main 
downside of using a deformed basis is the need to carefully study the 
convergence of calculations as a function of the basis deformation; see 
\cite{schunck2013density} for a discussion of the impact of basis 
truncation on HFB observables. In this section, we demonstrate that the convergence pattern of AMP calculations is
generally different from the one of the
underlying HFB calculations.

\begin{figure}[!ht]
\includegraphics[width=0.48\textwidth]{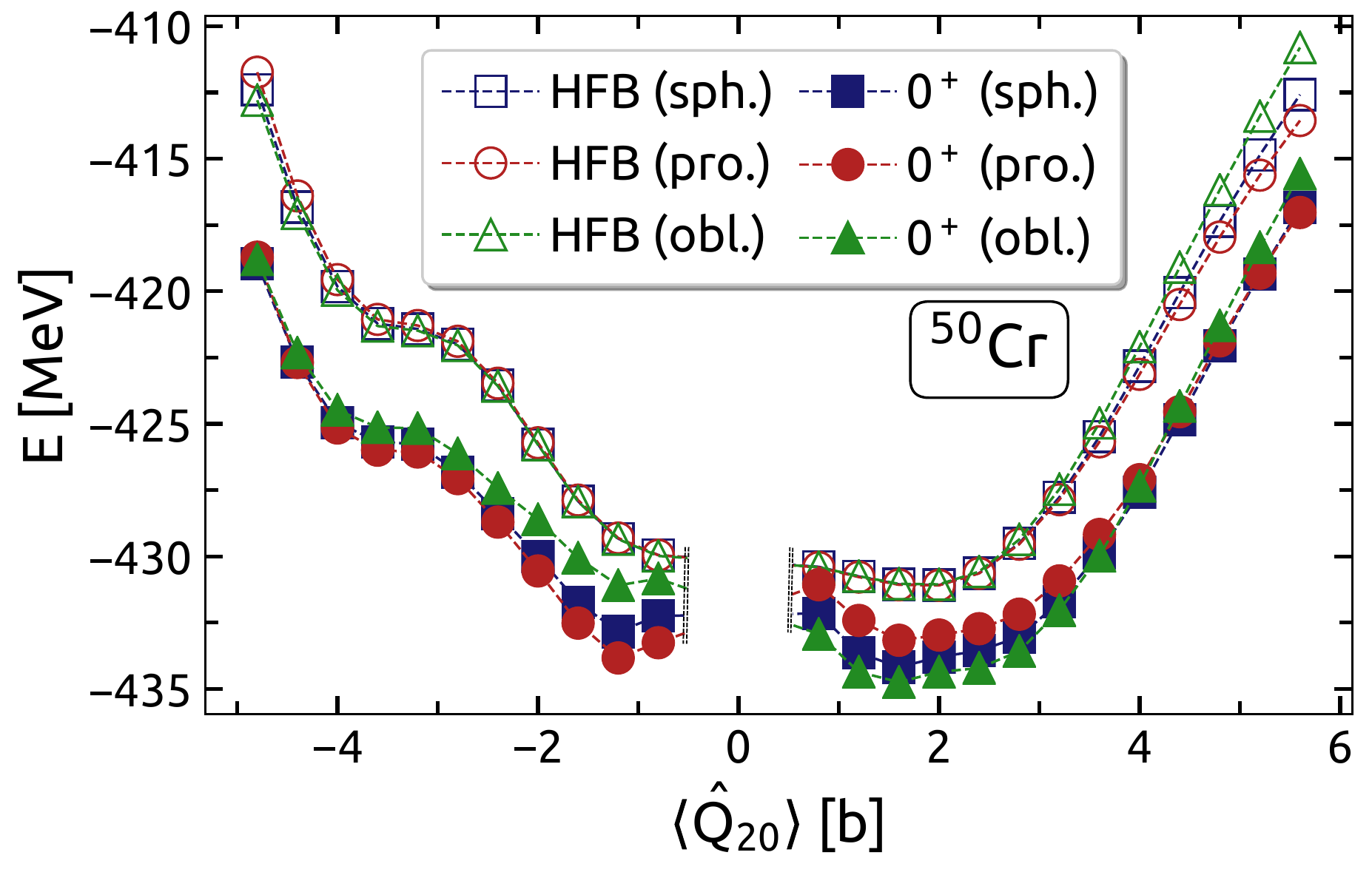}
\caption{Total HFB and $J^{p} = 0^{+}$ energy of $^{50}$Cr as a function of 
the constraint on the axial quadrupole moment $\braket{\hat{Q}_{20}}$. Blue 
curves with squares show results obtained with a spherical basis; red curves 
with circles show results obtained with a prolate-deformed basis of $\beta_2 = 0.1$; 
green curves with triangles show results 
obtained with an oblate-deformed basis 
of $\beta_2 = -0.1$. Plain 
symbols correspond to AMP results and open symbols to HFB ones; see text for 
additional details.
}
\label{fig:convergence_cr50}
\end{figure}

Fig.~\ref{fig:convergence_cr50} shows the HFB energy
and the AMP ($J^p = 0^+$)
energy in $^{50}$Cr
as a function of the
axial quadrupole moment $\braket{\hat{Q}_{20}}$ and 
obtained with three different HO bases:
the spherical ($\beta_2 \! = \! 0.0)$ basis,
the prolate-deformed ($\beta_2 = 0.1$) basis,
and the oblate-deformed ($\beta_2 \! = \! -0.1$) basis. $N_0 \! = \! 8$ HO shells were used in all three cases.
For configurations with moderate prolate
deformation,
the $0^+$ energies obey
$E_{J=0}(\beta_2 \! = \! -0.1) \! < \! E_{J=0}(\beta_2 = 0.0) \! < \! E_{J=0}(\beta_2 = 0.1)$.
The differences in HFB energies are much smaller,
but they obey the exact opposite rule:
$E_{\rm HFB}(\beta_2 \! = \! -0.1) 
\!> \! E_{\rm HFB}(\beta_2 \! = \! 0.0) \! > \! E_{\rm HFB}(\beta_2 \! = \! 0.1)$.
Interestingly, the pattern is reversed for 
configurations with moderate oblate deformation. For
them, the prolate-deformed basis gives the lowest $0^+$ energy and the oblate-deformed
basis gives the highest $0^+$
energy. In addition, the pattern is further modified as the deformation increases: for configurations with 
$\braket{\hat{Q}_{20}} \gtrapprox 5.4$~b the HFB and the $0^+$ energy follow the same ordering
and the lowest energies are obtained
with the prolate-deformed basis.

The observed difference in patterns may have two main origins:
\begin{itemize}
\item {\bf Numerical Precision.} For a prolate-deformed basis, the number of basis states 
along the $z$-axis of the reference frame, which coincides with the elongation axis 
of the HFB configuration, is larger than the number of
states along the perpendicular axis. Consequently, the prolate-deformed HFB 
configuration is numerically well described. However, the elongation axis of the rotated HFB
configuration is not anymore aligned with the 
$z$-axis of the reference frame. 
In fact, for $\beta \! = \! \pi/2$ it is aligned with the axis perpendicular to it -- where the number of basis 
states is lower. Rotated prolate-deformed configurations are thus described less precisely in a prolate-deformed basis. 
Moreover, the weight of each rotated configuration is
$\sin\beta\, d_{00}^{J}(\beta)$. For $J\!=\!0$, $d_{00}^0(\beta)=1$, and the
weight is simply $\sin\beta$. Consequently, the $\beta \! \approx \! \pi/2$ configurations, 
which are numerically less precise, have larger weights than the $\beta\approx 0$ 
configurations, which are numerically more precise. For $J>0$, the function 
$\sin\beta\, d_{00}^{J}(\beta)$ is not monotonous and this simple analysis does not hold anymore.
\item {\bf The Effect of the Rotation Matrix.} The rotation matrix 
[Eq.~\eqref{eq:rotation_matrix}] enters
the calculation of
overlaps [Eq.~\eqref{eq:overlap}]. Furthermore,
the overlaps enter the calculation of the 
norm overlap kernel $\mathcal{N}_{\bm{q}}^{J; p}$ and the Hamiltonian kernel $\mathcal{H}_{\bm{q}}^{J; p}$, both of which
are needed to calculate the AMP
energy [Eq.~\eqref{eq:projected_energy}]. 
However, the properties of the rotation matrix depend on the basis deformation. For example, the
determinant of the rotation matrix equals to $1$ in the spherical basis and
decreases rapidly
as the basis deformation increases. Without actually performing the calculations, 
it is not clear how the deformation {\it of the basis} impacts the rotation matrix, the subsequent kernels and, eventually, the AMP energy. 
\end{itemize}

\begin{figure}[!ht]
\includegraphics[width=0.48\textwidth]{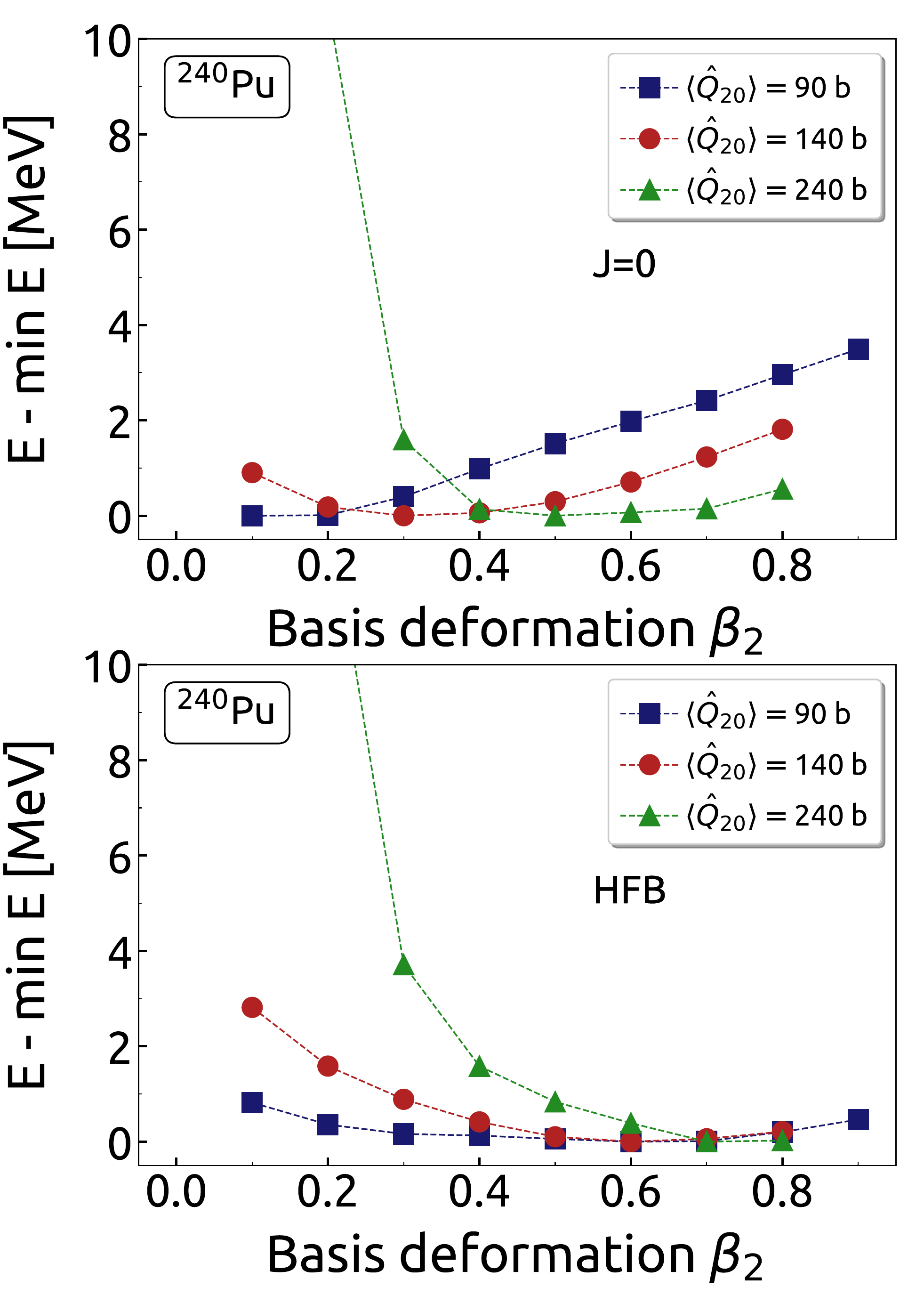}
\caption{The convergence of the HFB energy (bottom) and the AMP $0^+$ energy (top) as a function
of the basis deformation $\beta_2$ for three configurations along the fission path of $^{240}$Pu: $(Q_{20},Q_{30}) = (90 \, \mathrm{b},0 \,\mathrm{b}^{3/2})$, 
$(Q_{20},Q_{30}) = (140 \, \mathrm{b},12 \,\mathrm{b}^{3/2})$, 
and $(Q_{20},Q_{30}) = (240 \, \mathrm{b},25 \,\mathrm{b}^{3/2})$. All curves are normalized 
relative to their respective minima over the interval $\beta_2 \in [0,0.9]$; see text for 
additional details.
}
\label{fig:convergence_pu240}
\end{figure}

To get a better idea of the convergence pattern of AMP calculations as a 
function of the basis deformation, Fig.~\ref{fig:convergence_pu240} shows a semi-realistic example of the fission path of 
$^{240}$Pu. We considered three different configurations along the path: 
$(Q_{20},Q_{30}) = (90\, \mathrm{b},0\, \mathrm{b}^{3/2})$, 
$(Q_{20},Q_{30}) = (140\, \mathrm{b},12\, \mathrm{b}^{3/2})$, and 
$(Q_{20},Q_{30}) = (240\, \mathrm{b},25\, \mathrm{b}^{3/2})$. For each configuration, we computed the HFB solution in a basis characterized by 
$N_0^{\rm max} = 24$ HO shells and $\beta_2 \! = \! 0.0, 0.1, ..., 0.9$ deformation. In
addition, the basis was truncated
and only the lowest $N_{\rm states} = 1100$ states were retained. 
The spherical-equivalent oscillator length $b_0$ was {\it not} adjusted and was instead
fixed at $b_0 = 2.288$ fm. In other words, the
oscillator lengths $b_{z}$ and $b_{\perp}$ vary as  
a function of $\beta_2$ in such a way that the product $b_{z}b^{2}_{\perp} = 
b_0^{3}$ is constant.

The HFB convergence pattern (bottom panel) should be familiar to  the practitioners: very deformed 
configurations require (very) deformed bases. In our example, the lowest HFB energy 
is found for $\beta_2 \! = \! 0.6$ ($\braket{\hat{Q}_{20}} \! = \! 90$ b and $\braket{\hat{Q}_{20}} \! = \! 140$ b) 
and for $\beta_2 = 0.8$ ($\braket{\hat{Q}_{20}} = 240$ b). Note that, in principle, one should also 
adjust the oscillator frequency as a function of the deformation; see 
discussion in \cite{schunck2013density}. For very deformed configurations, the 
convergence pattern of the $0^+$ energy is qualitatively similar to 
the HFB pattern in the sense that the minimum is obtained for non-zero
$\beta_2$ values. However, these values are significantly smaller than in the HFB 
case. In fact, for the least-deformed configuration (which approximately corresponds 
to the fission isomer), the lowest $0^+$ energy is obtained for a nearly 
spherical basis. These results suggest that large-scale applications of AMP in 
a deformed basis should be accompanied by a careful study of the numerical 
convergence.


\subsubsection{Limitations of the Model}
\label{subsubsec:limitations}

The user should be aware of a number of
limitations of the novel symmetry restoration module, related to both 
the underlying physics and
the numerical implementation:

\begin{itemize}
\item {\bf Projection of the Eigenstates.} Some HFB
configurations are already eigenstates of an
operator related to the symmetry being
restored. For example, the spherical configuration
is an eigenstate of the angular momentum
operator with
the eigenvalue $J=0$. Similarly, configurations 
with vanishing odd multipole moments are
eigenstates of the parity operator with the
eigenvalue $p=+1$. Projecting these configurations
onto other eigenvalues ($J\!=\!1,2, ...$ for the former
and $p\!=\!-1$ for the latter) will yield non-physical
results. In practice, one should be cautious because numerical issues
can occur already for configurations that
are sufficiently close to being eigenstates.
\item{\bf Invertibility of the Rotation Matrix.} The inverse and the determinant of the rotation matrix
enter our calculations explicitly.
However, as the size and the deformation of the
basis increase, the determinant drops rapidly and the matrix can become
numerically non-invertible for some
rotational angles close to $\beta = \pi/2$.
These angles are then disregarded in AMP,
under the assumption that the corresponding
overlaps are negligible. This assumption is
justified
for very deformed configurations, but it can
break down for configurations with moderate
or small deformations. Consequently, caution
is advised when calculating moderately
deformed configurations with deformed bases.
In particular, the description of near-spherical configurations 
with deformed bases is imprecise
and should therefore be avoided.

\item{\bf Spuriosity of Projected Energies.} The Hamiltonian kernel is formally not
well-defined for EDFs that are density-dependent or omit parts of the interaction. In the worst case scenario, this can lead to sizeable finite steps
and even divergences in projected energies. Such spuriosities were 
abundantly reported in PNP \cite{anguiano2001particle,
dobaczewski2007particlenumber, duguet2009particlenumber,
bender2009particlenumber}, while
AMP in even-even nuclei seems
to remain issue-free \cite{egido2016stateoftheart}.
In many practical implementations, however, the scale of these spuriosities
is smaller than the errors
due to the various numerical limitations. Nevertheless, as the quest
for spuriosity-free EDFs is under way, the 
user should remain aware of this 
formal limitation.
\end{itemize}


\subsection{Exact Coulomb}
\label{subsec:coulomb}

We tested our implementation of the "exact" Coulomb calculation by comparing 
results obtained with the new version of \pr{hfbtho} and with the 
Gogny code used in \cite{schunck2008continuum,schunck2008nuclear}. In 
the latter, all contributions of the Coulomb interaction (direct, exchange, 
and pairing) are computed exactly thanks to the properties of the spherical 
HO basis. 

\begin{figure}[!ht]
\includegraphics[width=0.49\textwidth]{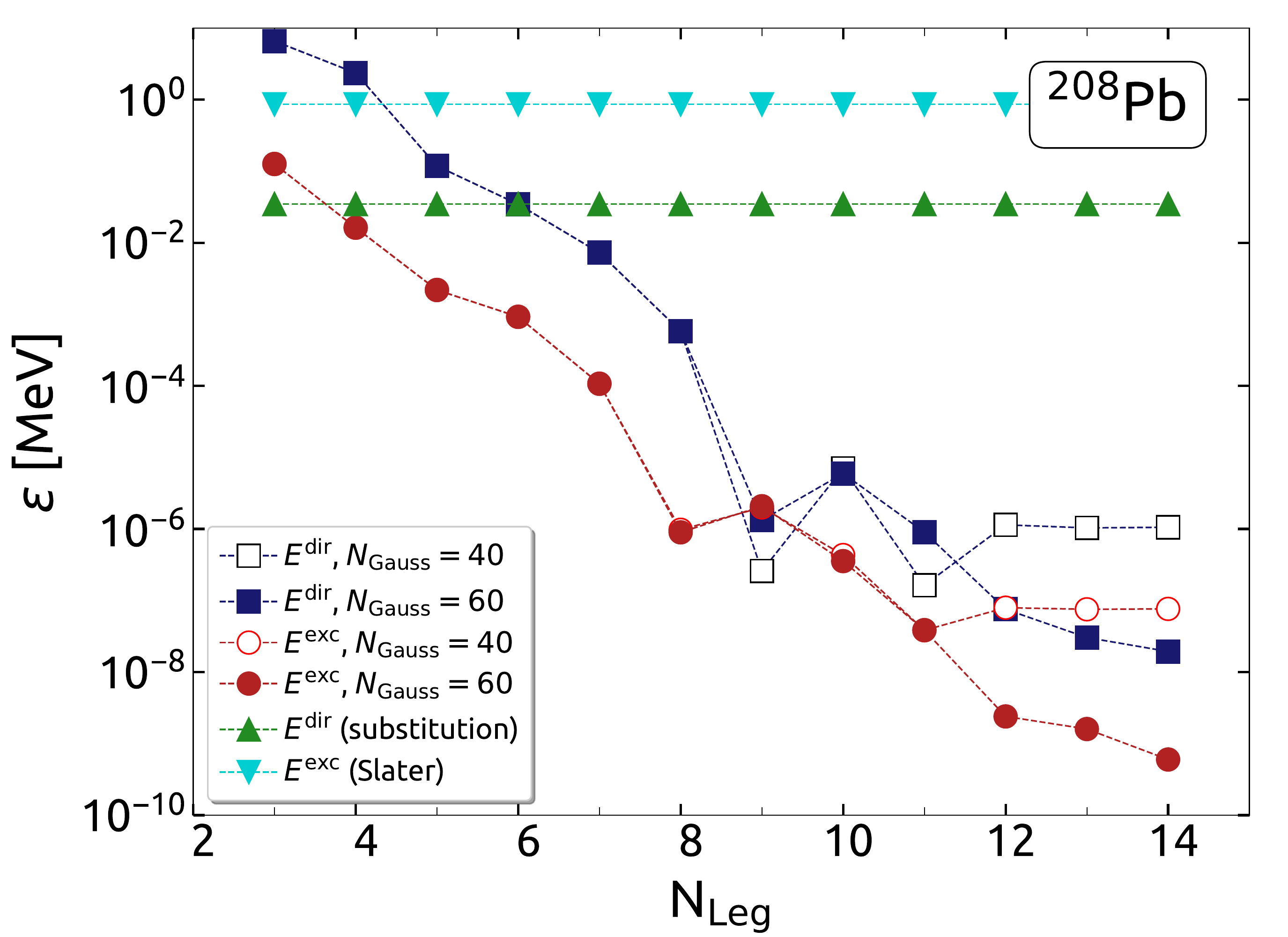}
\caption{The absolute error (in MeV) of the Gaussian expansion of the Coulomb 
potential as a function of Gauss-Legendre quadrature points, i.e., the number 
of Gaussians approximating $1/r$; see Eq.~\eqref{eq:coulomb}.}
\label{fig:coulomb}
\end{figure}

For numerical comparison, we consider the $^{208}$Pb nucleus and use the D1S 
Gogny EDF. Furthermore, we disregard the two-body center-of-mass correction and neglect the 
Coulomb contribution to pairing. Calculations are performed in a spherical HO basis
with $N_0 = 12$ shells and the oscillator length
$b_0 = 2.5$ fm. They
were converged up to $10^{-12}$. Fig. \ref{fig:coulomb} shows the 
absolute error $\varepsilon = |E^{X}_{ {\text{\hfbtho}}} - E^{X}_{\rm Gogny}|$ as a function of the number of
Gauss-Legendre quadrature points $N_{\rm Leg}$. 
Here, $X$ stands for either the direct or the exchange contribution to the 
Coulomb energy, and the subscripts "\hfbtho" and "Gogny" refer to the \pr{hfbtho} 
{\codeversion} and the spherical Gogny code, respectively.

For $N_{\rm Gauss} \! = \! 60$ points in both the Gauss-Hermite and Gauss-Laguerre integrations 
(the full lines), the expansion of the Coulomb 
potential onto Gaussians converges nicely to the exact value. In particular, at $N_{\rm{Leg}} = 14$, the difference is $20$ meV and 
$1$ meV for the direct and the exchange term, respectively. If 
the number of quadrature points is reduced to $N_{\rm Gauss} = 40$ (the dashed lines), we observe a 
saturation of convergence at about $1$ eV (direct) and $80$ meV (exchange) at 
$N_{\rm Leg} = 14$. For comparison, we also show the results of 
the "standard" prescription for the direct term, which is based on the 
substitution method in a box of size $L = 50$ fm with $80$ Gauss-Legendre 
quadrature points; see discussion in \cite{stoitsov2013axially}, and for the 
exchange term, which is computed at the Slater approximation.


\section{Input data file}
\label{sec:input}

The input data file format remains similar to version 3.00 and only contains 
one additional namelist.


\subsection{Sample input file}
\label{subsec:sample}

\begin{verbatim}
&HFBTHO_GENERAL
 number_of_shells = 10, 
 oscillator_length = -1.0, 
 basis_deformation = 0.0, 
 proton_number = 24, neutron_number = 26,
 type_of_calculation = 1 /
 &HFBTHO_INITIAL
 beta2_deformation = 0.0, 
 beta3_deformation = 0.0, 
 beta4_deformation = 0.0 /
&HFBTHO_ITERATIONS
 number_iterations = 100, accuracy = 1.E-5, 
 restart_file = -1 /
&HFBTHO_FUNCTIONAL
 functional = 'SLY4', 
 add_initial_pairing = F, 
 type_of_coulomb = 2 /
&HFBTHO_PAIRING
 user_pairing = F, 
 vpair_n = -300.0, vpair_p = -300.0,
 pairing_cutoff = 60.0,
 pairing_feature = 0.5 /
&HFBTHO_CONSTRAINTS
 lambda_values = 1, 2, 3, 4, 5, 6, 7, 8,
 lambda_active = 0, 0, 0, 0, 0, 0, 0, 0,
 expectation_values = 0.0, 0.0, 0.0, 0.0,
 0.0, 0.0, 0.0, 0.0 /
&HFBTHO_BLOCKING
 proton_blocking = 0, 0, 0, 0, 0, 
 neutron_blocking = 0, 0, 0, 0, 0 /
&HFBTHO_PROJECTION
 switch_to_THO = 0, 
 projection_is_on = 0, gauge_points = 1,
 delta_Z = 0, delta_N = 0 /
&HFBTHO_TEMPERATURE
 set_temperature = F, temperature = 0.0 /
&HFBTHO_FEATURES
 collective_inertia = F, 
 fission_fragments = F, 
 pairing_regularization = F, 
 localization_functions = F /
&HFBTHO_NECK
 set_neck_constrain = F, neck_value = 0.5 /
&HFBTHO_DEBUG
 number_Gauss = 40, number_Laguerre = 40, 
 number_Legendre = 80, 
 compatibility_HFODD = F, 
 number_states = 500,  
 force_parity = T, print_time = 0 /
&HFBTHO_RESTORATION
 PNP_is_on = 0, number_of_gauge_points = 1, 
 delta_neutrons = 0, delta_protons = 0,
 AMP_is_on = 0, 
 number_of_rotational_angles = 1, 
 maximal_angular_momentum = 0 /
\end{verbatim}


\subsection{Description of input data}
\label{subsec:description}

We now define the new or updated inputs introduced in version {\codeversion}.

\key{HFBTHO\_FUNCTIONAL}

\noindent$\bullet$ {\tt type\_of\_coulomb = 2}: Logical switch that defines the 
treatment of the Coulomb potential. In previous versions, this switch could 
only take values $0$ (no Coulomb), $1$ (direct contribution only) or $2$ (direct and exchange contribution with the Slater approximation). In the current 
version, the following new options are also available: 
\begin{enumerate}
\item[-1:] direct Coulomb only by sum of $N_c$ Gaussians;
\item[-2:] direct Coulomb by the substitution method, exchange Coulomb by sum 
of $N_c$ Gaussians;
\item[-3:] direct Coulomb by sum of $N_c$ Gaussians, exchange Coulomb with the 
Slater approximation;
\item[-4:] direct and exchange Coulomb by sum of $N_c$ Gaussians;
\item[-5:] direct, exchange, and pairing Coulomb by sum of $N_c$ Gaussians.
\end{enumerate}
Here, $N_c$ is the number of Gaussians in \eqref{eq:coulomb}. It is stored in 
the UNEDF module variable {\tt n\_g\_coul} and is preset at {\tt n\_g\_coul=9} 
in the file {\tt hfbtho\_unedf.f90}. There is no option to change this number 
directly in the input file. 
Default:~2. \\

\key{HFBTHO\_RESTORATION}

\noindent$\bullet$ {\tt PNP\_is\_on = 0}: Logical switch that activates the particle
number projection in the quasiparticle basis. 
When set to $1$ the mixed density prescription
is used and when set to $2$ the projected
density prescription is used (see Sections \ref{subsubsec:hamiltonian_kernel} and \ref{subsubsec:prescriptions}). This 
option is different from the old {\tt projection\_is\_on} switch in the
{\tt HFBTHO\_PROJECTION} namelist, which activates PNP
with the mixed density prescription in the canonical basis. For 
an infinite quasiparticle cutoff, 
the two mixed density
prescription options should give the same result. This option is incompatible 
with: finite-temperature, THO basis, and blocking calculations.
Default:~0; \\

\noindent$\bullet$ {\tt number\_of\_gauge\_points = 1}: Number of gauge angles 
$N_{\varphi}$ for particle number projection. The same number $N_{\varphi}$ is 
used for protons and neutrons. 
Default: 1; \\

\noindent$\bullet$ {\tt delta\_neutrons = 0}: Value of the shift in neutron 
number $\delta N$. In the case of PNP, one can project on all even neutron numbers 
in the interval $[N_0 -\delta N, N_0 + \delta N  ]$, where $N_0$ 
is the number of neutrons of the considered nucleus (even only for PNP). 
Default: 0; \\

\noindent$\bullet$ {\tt delta\_protons = 0}: Value of the shift in proton 
number $\delta Z$. In the case of PNP, one can project on all even proton numbers in the interval $[Z_0 -\delta Z, Z_0 + \delta Z ]$, where $Z_0$ 
is the number of protons of the considered nucleus (even only for PNP). 
Default: 0; \\

\noindent$\bullet$ {\tt AMP\_is\_on = 0}: Logical switch that activates (if 
equal to 1) the restoration of angular momentum $J$
and parity $p$. 
This option can be combined with PNP to carry out a simultaneous projection on 
$N$, $Z$, $J$, and $p$. It is incompatible 
with: finite-temperature, THO basis, and blocking calculations.
Default: 0; \\

\noindent$\bullet$ {\tt number\_of\_rotational\_angles = 1}: Number of 
rotational angles $N_{\beta}$ use for AMP. Internally, the code will readjust 
$N_{\beta}$ if reflection symmetry is enforced. In such a case, the 
program will compute either $N_{\beta}/2$ ($N_{\beta}$ even) or 
$(N_{\beta}+1)/2$ ($N_{\beta}$ odd) rotational angles (see 
Section~\ref{subsubsec:symmetries}).
Default: 1; \\

\noindent$\bullet$ {\tt maximal\_angular\_momentum = 0}: Maximum value of the 
angular momentum $J_{\rm max}$. In the case of AMP, all even values of $J$ in 
$[0,J_{\rm max}]$ (parity conserved) or all values $J$ in $[0,J_{\rm max}]$.
Default: 0. \\


\section{Program \textsc{hfbtho}}
\label{sec:program}


\subsection{Structure of the code}
\label{subsec:code}

Compared with version 3.00, we have substantially increased the modularization 
of the source code since the number of modules increased from 18 to 25. The 
code is organized as follows:
\begin{itemize}
\item {\tt hfbtho\_bessel.f90}: defines the modified Bessel functions of order 
0 and 1;
\item {\tt hfbtho\_canonical.f90}: defines the canonical basis of the HFB 
theory;
\item {\tt hfbtho\_collective.f90}: computes the ATDHF and GCM collective 
inertia tensor and zero-point energy correction in the perturbative cranking 
approximation; see \cite{perez2017axially} and references therein;
\item {\tt hfbtho\_elliptic\_integrals.f90}: defines complete elliptic integral 
of the second kind used for the Coulomb potential;
\item {\tt hfbtho\_fission.f90}: computes the charge, mass, and axial multipole 
moments of fission fragments and the value of the Gaussian neck operator;
\item {\tt hfbtho\_gauss.f90}: defines the quadrature meshes: Gauss-Hermite, 
Gauss-Laguerre, and Gauss-Legendre;
\item {\tt hfbtho\_gogny.f90}: computes the matrix elements of the Gogny force 
as well as the corresponding mean field and pairing field;
\item {\tt hfbtho\_io.f90}: contains a collection of routines handling inputs 
and outputs;
\item {\tt hfbtho\_large\_scale.f90}: contains a collection of routines for mass 
table, drip lines, or potential energy surface calculations, as well as for the 
parallelization of single HFB calculations;
\item {\tt hfbtho\_library.f90}: provides the definition of the main routine 
{\tt Main\_Program()} that launches complete {\hfbtho} calculations: 
stand-alone, mass tables, drip lines, or potential energy surfaces;
\item {\tt hfbtho\_lipkin.f90}: calculates the Lipkin-Nogami correction, 
including the $\lambda_2$ parameters, densities, and energies;
\item {\tt hfbtho\_localization.f90}: computes spatial localization functions;
\item {\tt hfbtho\_main.f90}: calls the {\tt Main\_Program()} routine;
\item {\tt hfbtho\_math.f90}: contains a collection of general-use mathematical 
routines;
\item {\tt hfbtho\_multipole\_moments.f90}: computes the expectation value and 
matrix elements of axial multipole moments; 
\item {\tt hfbtho\_pnp.f90}: implements particle
number projection in the canonical basis;
\item {\tt hfbtho\_projections.f90}: implements the angular momentum, particle 
number, and parity projection in the quasiparticle basis;
\item {\tt hfbtho\_read\_functional.f90}: contains a collection of routines to 
read the parameters of the EDF from a file;
\item {\tt hfbtho\_solver.f90}: solves the self-consistent iterations of the 
HFB theory;
\item {\tt hfbtho\_storage.f90}: contains an interface to the QRPA pnFAM code; 
see \cite{ney2020global} and references therein;
\item {\tt hfbtho\_tho.f90}: defines the transformed harmonic oscillator basis; 
see \cite{stoitsov2005axially} and references therein;
\item {\tt hfbtho\_unedf.f90}: defines parameterizations of the Skyrme and 
Gogny functionals, and computes density-dependent coupling constants and fields 
of generalized Skyrme energy functionals;
\item {\tt hfbtho\_utilities.f90}: defines the integer and real types used 
throughout the code, as well as various numerical constants;
\item {\tt hfbtho\_variables.f90}: contains list of global variables used 
throughout the code;
\item {\tt hfbtho\_version.f90}: version number (currently git commit number of 
the previous commit) and history of previous versions.
\end{itemize}

The programming language of most of the code is now Fortran 2003. The code 
\pr{{\hfbtho}} requires an implementation of the BLAS and LAPACK libraries to
function correctly. Shared memory parallelism is available via OpenMP pragmas. 

This version comes with a built-in Doxygen documentation. To benefit from this 
feature, the user should install the doxygen software available at 
\url{www.doxygen.org}. The documentation is built by typing
\begin{center}
\begin{tabular}{p{0.35\textwidth}}
{\tt make doc }
\end{tabular}
\end{center}
By default, Doxygen generates only an on-line HTML documentation. The main 
page is located in the source directory at {\tt ./src/doc/html/index.html}. A 
PDF documentation can also be generated by going into {\tt ./doc/latex} and 
typing
\begin{center}
\begin{tabular}{p{0.35\textwidth}}
{\tt make}
\end{tabular}
\end{center}
The PDF file is named {\tt refman.pdf}.


\subsection{Running the code}
\label{subsec:run}

The program ships with a Makefile that is preset for a number of Fortran 
compilers. The user should choose the compiler and set the path for the BLAS 
and LAPACK libraries. In version {\codeversion} of the code, we have simplified 
the call sequence of {\hfbtho}. Assuming an executable named {\tt hfbtho\_main} 
and a Linux system, execution is started by typing
\begin{center}
\begin{tabular}{p{0.35\textwidth}}
{\tt ./hfbtho\_main [input\_file\_name] }
\end{tabular}
\end{center}
where {\tt [input\_file\_name]} is an optional name of the {\hfbtho} input file 
that contains all the Namelists. If none is given, the code will attempt to 
read the file with the generic name {\tt hfbtho\_NAMELIST.dat} in the current 
directory. The code will also automatically generate two ASCII output files: a 
compact one called {\tt hfbtho.out} and a more extended one called 
{\tt thoout.dat}. Finally, the code generates a binary file named 
{\tt hfbtho\_output.hel} that is used to restart calculations.

HFB calculations are greatly 
accelerated when OpenMP multi-threading is activated. However, the user 
should keep in mind that this requires setting additional environment 
variables. In Linux/Unix machines, the default stack size is not large 
enough to run the code and must be increased. This can be achieved by 
instructions such as

\begin{table}[!h]
\centering
\begin{tabular}{p{0.35\textwidth}}
\noindent {\tt ulimit -s unlimited}\\
\noindent {\tt export OMP\_STACKSIZE=32M}
\end{tabular}
\end{table}

The value of {\tt ulimit}  defines the amount of stack size for the main OpenMP 
thread. OpenMP supports control over the stack size limit of all additional 
threads via the environment variable {\tt OMP\_STACKSIZE}. The value given 
above should be sufficient for all applications. Note that this value does not 
affect the stack size of the main thread set by {\tt ulimit}. For completeness, 
note that the GNU OpenMP run-time ({\tt libgomp}) recognizes the non-standard 
environment variable {\tt GOMP\_STACKSIZE}. If set, it overrides the value of 
{\tt OMP\_STACKSIZE}. Finally, the Intel OpenMP run-time library also 
recognizes the non-standard environment variable {\tt KMP\_STACKSIZE}. If set, 
it overrides the value of both {\tt OMP\_STACKSIZE} and {\tt GOMP\_STACKSIZE}.


\section*{Acknowledgments}
\label{sec:acknowledgments}

Support for this work was partly provided through Scientific Discovery
through Advanced Computing (SciDAC) program funded by U.S. Department of
Energy, Office of Science, Advanced Scientific Computing Research and
Nuclear Physics. It was partly performed under the auspices of the US
Department of Energy by the Lawrence Livermore National Laboratory under
Contract DE-AC52-07NA27344 (code release number: LLNL-CODE-826901, document
release number: LLNL-JRNL-827553).
This work has been supported in part by the QuantiXLie
Centre of Excellence, a project cofinanced by the Croatian
Government and European Union through the European
Regional Development Fund - The Competitiveness and
Cohesion Operational Programme (KK.01.1.1.01.0004).  Computing support for this work came from 
the Lawrence Livermore National Laboratory (LLNL) Institutional Computing Grand 
Challenge program.

\appendix
\section{Densities and Currents in the Coordinate-Space Representation}
\label{sec:app_densities}
Taking into account
the block structure of the density matrix
in the $y$-simplex basis 
[cf. Eq. \eqref{eq:density_matrix}], we
can write
\begin{align}
\begin{split}
\rho^{(\tau)}(\bm{r}\sigma,\bm{r'}\sigma') =& \sum_{\alpha \gamma} 
\rho_{\alpha \gamma}^{(\tau)++} \Phi_{\gamma}^{s=+\di *}(\bm{r'}\sigma')
\Phi_{\alpha}^{s=+\di}(\bm{r}\sigma)\\
+& \sum_{\alpha \gamma} \rho_{\alpha \gamma}^{(\tau)--} 
\Phi_{\gamma}^{s=-\di *}(\bm{r'}\sigma')
\Phi_{\alpha}^{s=-\di}(\bm{r}\sigma),
\end{split}
\label{eq:density_simplex_expansion}
\end{align}
where the sums run over
HO basis states $\alpha$ and $\gamma$, while 
$\Phi_{\gamma}^{s=+\di}(\bm{r}\sigma)$ and 
$\Phi_{\gamma}^{s=-\di}(\bm{r}\sigma)$ are the 
coordinate space
representations of the eigenstates of the
$y$-simplex operator 
[cf. Eqs. \eqref{eq:simplexbasis}
and \eqref{eq:simplexoperator}]
\begin{subequations}
\begin{align}
\Phi_{\gamma}^{s=+\di}(\bm{r}\sigma) &= 
\frac{1}{\sqrt{4\pi}}\psi_{n_z^{\alpha}}(z)
\psi_{n_{\perp}^{\alpha}}^{|\Lambda^{\alpha}|}(r_{\perp}) 
\nonumber \\ & \times \Big[\di e^{\di \Lambda^\alpha \phi} 
\chi_{+\frac{1}{2}}(\sigma) + e^{-\di \Lambda^\alpha \phi} 
\chi_{-\frac{1}{2}}(\sigma) \Big], \\
\Phi_{\gamma}^{s=-\di}(\bm{r}\sigma) &= 
\frac{1}{\sqrt{4\pi}}\psi_{n_z^{\alpha}}(z)
\psi_{n_{\perp}^{\alpha}}^{|\Lambda^{\alpha}|}(r_{\perp})
\nonumber \\ & \times \Big[e^{\di \Lambda^\alpha \phi} 
\chi_{+\frac{1}{2}}(\sigma) + \di e^{-\di \Lambda^\alpha \phi}
\chi_{-\frac{1}{2}}(\sigma) \Big].
\end{align}
\end{subequations}
Components of the HO eigenfunctions $\psi_{n_z^{\alpha}}(z)$ and $
\psi_{n_{\perp}^{\alpha}}^{|\Lambda^{\alpha}|}(r_{\perp})$
are defined in \cite{stoitsov2005axially} and
$\chi_{\pm \frac{1}{2}}(\sigma)$ are the
eigenstates of the $z$-component
of the spin operator. Note that in
Eq. \eqref{eq:density_simplex_expansion} the
dependence on $\bm{x}^{(\tau)}$ and $\bm{q}$ was
dropped for compactness in both $\rho_{\bm{q}}^{(\tau)}(\bm{r}\sigma,\bm{r'}\sigma'; \bm{x}^{(\tau)})$ on the left and $\rho_{\bm{q}, \alpha \gamma}^{(\tau)++}(\bm{x}^{(\tau)})$,
$\rho_{\bm{q}, \alpha \gamma}^{(\tau)--}(\bm{x}^{(\tau)})$ on the right.

The auxiliary local densities \eqref{eq:particle_density}-\eqref{eq:spin_current_density} 
can then be
calculated from Eq. \eqref{eq:density_simplex_expansion} as
\begin{subequations}
\begin{align}
\rho^{(\tau)}(\bm{r})&=\sum_{\alpha \gamma}
\rho_{\alpha \gamma, +}^{(\tau)} \mathcal{F}^1_{\alpha \gamma}(r_{\perp},z)
\cos\Big[(\Lambda^{\!\alpha}\!-\!\Lambda^{\beta})\phi\Big], \! \\
s_{r_\perp}^{(\tau)}(\bm{r})&=\!-\sum_{\alpha \gamma}
\rho_{\alpha \gamma, -}^{(\tau)} \mathcal{F}^1_{\alpha \gamma}(r_{\perp},\!z)
\sin\Big[(\Lambda^{\!\alpha}\!+\!\Lambda^{\beta}\!+\!1)\phi\Big], \! \\ 
s_{\phi}^{(\tau)}(\bm{r})&=\!-\sum_{\alpha \gamma}
\rho_{\alpha \gamma, -}^{(\tau)} \mathcal{F}^1_{\alpha \gamma}(r_{\perp},\!z)
\cos\Big[(\Lambda^{\!\alpha}\!\!+\!\Lambda^{\beta}\!+\!1)\phi\Big], \! \\
s_{z}^{(\tau)}(\bm{r})&=\di\sum_{\alpha \gamma}
\rho_{\alpha \gamma, +}^{(\tau)} \mathcal{F}^1_{\alpha \gamma}(r_{\perp},z)
\sin\Big[(\Lambda^{\!\alpha}\!-\!\Lambda^{\beta})\phi\Big], \! \\
\tau^{(\tau)}(\bm{r})&=\sum_{\alpha \gamma}
\rho_{\alpha \gamma, +}^{(\tau)} \mathcal{F}^2_{\alpha \gamma}(r_{\perp},z)
\cos\Big[(\Lambda^{\!\alpha}\!-\!\Lambda^{\beta})\phi\Big], \! \\
T_{r_\perp}^{(\tau)}(\bm{r})&=\!-\sum_{\alpha \gamma}
\rho_{\alpha \gamma, -}^{(\tau)} \mathcal{F}^3_{\alpha \gamma}(r_{\perp},z)
\sin\Big[(\Lambda^{\!\alpha}\!\!+\!\Lambda^{\beta}\!+\!1)\phi\Big], \!  \\
T_{\phi}^{(\tau)}(\bm{r})&=\!-\sum_{\alpha \gamma}
\rho_{\alpha \gamma, -}^{(\tau)} \mathcal{F}^3_{\alpha \gamma}(r_{\perp},\!z)
\cos\Big[(\Lambda^{\!\alpha}\!\!+\!\Lambda^{\beta}\!+\!1)\phi\Big], \! \\
T_{z}^{(\tau)}(\bm{r})&=\di\sum_{\alpha \gamma}
\rho_{\alpha \gamma, +}^{(\tau)} \mathcal{F}^2_{\alpha \gamma}(r_{\perp},z)
\sin\Big[(\Lambda^{\!\alpha}\!-\!\Lambda^{\beta})\phi\Big], \! \\
j_{r_\perp}^{(\tau)}(\bm{r})&=\frac{1}{2\di}\sum_{\alpha \gamma}
\rho_{\alpha \gamma, +}^{(\tau)} \mathcal{F}^4_{\alpha \gamma}(r_{\perp},z)
\cos\Big[(\Lambda^{\!\alpha}\!-\!\Lambda^{\beta})\phi\Big], \! \\
j_{\phi}^{(\tau)}(\bm{r})&=\frac{1}{2\di}\sum_{\alpha \gamma}
\rho_{\alpha \gamma, +}^{(\tau)} \mathcal{F}^5_{\alpha \gamma}(r_{\perp},z)
\sin\Big[(\Lambda^{\beta}\!-\!\Lambda^{\!\alpha})\phi\Big], \! \\
j_{z}^{(\tau)}(\bm{r})&=\frac{1}{2\di}\sum_{\alpha \gamma}
\rho_{\alpha \gamma, +}^{(\tau)} \mathcal{F}^6_{\alpha \gamma}(r_{\perp},z)
\cos\Big[(\Lambda^{\!\alpha}\!-\!\Lambda^{\beta})\phi\Big], \! \\
J_{r_\perp r_\perp}^{(\tau)}(\bm{r})&=\di \sum_{\alpha \gamma}
\rho_{\alpha \gamma, -}^{(\tau)} \mathcal{F}^4_{\alpha \gamma}(r_{\perp},z)
\sin\Big[(\Lambda^{\!\alpha}\!+\!\Lambda^{\beta}\!+\!1)\phi\Big], \! \\
J_{r_\perp \phi}^{(\tau)}(\bm{r})&=\!\di \sum_{\alpha \gamma}
\rho_{\alpha \gamma, -}^{(\tau)} \mathcal{F}^4_{\alpha \gamma}(r_{\perp},\!z)
\cos\Big[(\Lambda^{\!\alpha}\!\!+\!\Lambda^{\beta} \!+ \!1)\phi\Big], \! \\
J_{r_\perp z}^{(\tau)}(\bm{r})&= \sum_{\alpha \gamma}
\rho_{\alpha \gamma, +}^{(\tau)} \mathcal{F}^4_{\alpha \gamma}(r_{\perp},z)
\sin\Big[(\Lambda^{\!\alpha}\!-\!\Lambda^{\beta})\phi\Big], \! \\
J_{\phi r_\perp}^{(\tau)}(\bm{r})&=\! \di \sum_{\alpha \gamma}
\rho_{\alpha \gamma, -}^{(\tau)} \mathcal{F}^7_{\alpha \gamma}(r_{\perp},\!z)
\cos\Big[(\Lambda^{\!\alpha}\!+\!\Lambda^{\beta}\!+\!1)\phi\Big], \! \\
J_{\phi \phi}^{(\tau)}(\bm{r})&= \!-\di \! \sum_{\alpha \gamma}
\rho_{\alpha \gamma, -}^{(\tau)} \mathcal{F}^7_{\alpha \gamma}(r_{\perp},\!z)
\sin\Big[(\Lambda^{\!\alpha}\!\!+\!\Lambda^{\beta}\!+\!1)\phi\Big],\! \\
J_{\phi z}^{(\tau)}(\bm{r})&= \sum_{\alpha \gamma}
\rho_{\alpha \gamma, +}^{(\tau)} \mathcal{F}^5_{\alpha \gamma}(r_{\perp},z)
\cos\Big[(\Lambda^{\!\alpha}\!-\!\Lambda^{\beta})\phi\Big], \! \\
J_{z r_\perp}^{(\tau)}(\bm{r})&= \di \sum_{\alpha \gamma}
\rho_{\alpha \gamma, -}^{(\tau)} \mathcal{F}^6_{\alpha \gamma}(r_{\perp},z)
\sin\Big[(\Lambda^{\!\alpha}\!+\!\Lambda^{\beta}\!+\!1)\phi\Big], \! \\
J_{z \phi}^{(\tau)}(\bm{r})&= \! \di \sum_{\alpha \gamma}
\rho_{\alpha \gamma, -}^{(\tau)} \mathcal{F}^6_{\alpha \gamma}(r_{\perp},z)
\cos\Big[(\Lambda^{\! \alpha}\!+\!\Lambda^{\beta}\!+\!1)\phi\Big], \! \\
J_{z z}^{(\tau)}(\bm{r})&= \sum_{\alpha \gamma}
\rho_{\alpha \gamma, +}^{(\tau)} \mathcal{F}^6_{\alpha \gamma}(r_{\perp},z)
\sin\Big[(\Lambda^{\!\alpha}\!-\!\Lambda^{\beta})\phi\Big].
\end{align}
\end{subequations}
Here, we have introduced a shorthand notation
for density matrices
\begin{subequations}
\begin{align}
\rho_{\alpha \gamma, +}^{(\tau)} &= \frac{1}{2\pi}\Big( 
\rho_{\alpha \gamma}^{(\tau) ++} \!+\! \rho_{\alpha \gamma}^{(\tau) --} \Big), \\
\rho_{\alpha \gamma, -}^{(\tau)} &= \frac{1}{2\pi}\Big( 
\rho_{\alpha \gamma}^{(\tau) ++} \!-\! \rho_{\alpha \gamma}^{(\tau) --} \Big),
\end{align}
\end{subequations}
as well as for the coordinate-dependent factors
\begin{subequations}
\begin{align}
\mathcal{F}^1_{\alpha \gamma}(r_{\perp},z) &= 
\psi_{n_z^{\alpha}}(z) \psi_{n_{\perp}^{\alpha}}^{|\Lambda^{\!\alpha}|}(r_{\perp})
\psi_{n_z^{\beta}}(z) \psi_{n_{\perp}^{\beta}}^{|\Lambda^{\beta}|}(r_{\perp}), \\
\mathcal{F}^2_{\alpha \gamma}(r_{\perp},z) &= 
\psi_{n_z^{\alpha}}(z) \Big( \partial_{r_\perp} 
\psi_{n_{\perp}^{\alpha}}^{|\Lambda^{\!\alpha}|}(r_{\perp})\Big)
\psi_{n_z^{\beta}}(z) \Big(\partial_{r_\perp} 
\psi_{n_{\perp}^{\beta}}^{|\Lambda^{\beta}|}(r_{\perp})\Big) \nonumber \\
 & + \frac{\Lambda^{\!\alpha} \Lambda^\beta}{r^2_\perp} 
\mathcal{F}^1_{\alpha \gamma}(r_{\perp},z) \\
&+ \Big( \partial_z \psi_{n_z^{\alpha}}(z)\Big) 
\psi_{n_{\perp}^{\alpha}}^{|\Lambda^{\! \alpha}|}(r_{\perp})
\Big( \partial_z \psi_{n_z^{\beta}}(z)\Big) 
\psi_{n_{\perp}^{\beta}}^{|\Lambda^{\beta}|}(r_{\perp}), \nonumber \\
\mathcal{F}^3_{\alpha \gamma}(r_{\perp},z) &= 
\psi_{n_z^{\alpha}}(z) \Big( \partial_{r_\perp} 
\psi_{n_{\perp}^{\alpha}}^{|\Lambda^{\!\alpha}|}(r_{\perp})\Big)
\psi_{n_z^{\beta}}(z) \Big(\partial_{r_\perp} 
\psi_{n_{\perp}^{\beta}}^{|\Lambda^{\beta}|}(r_{\perp})\Big) \nonumber \\
 & - \frac{\Lambda^{\!\alpha} \Lambda^\beta}{r^2_\perp} 
 \psi_{n_z^{\alpha}}(z) \psi_{n_{\perp}^{\alpha}}^{|\Lambda^{\! \alpha}|}(r_{\perp})
\psi_{n_z^{\beta}}(z) \psi_{n_{\perp}^{\beta}}^{|\Lambda^{\beta}|}(r_{\perp}) \\
&+ \Big( \partial_z \psi_{n_z^{\alpha}}(z)\Big) 
\psi_{n_{\perp}^{\alpha}}^{|\Lambda^{\! \alpha}|}(r_{\perp})
\Big( \partial_z \psi_{n_z^{\beta}}(z)\Big) 
\psi_{n_{\perp}^{\beta}}^{|\Lambda^{\beta}|}(r_{\perp}), \nonumber \\
\mathcal{F}^4_{\alpha \gamma}(r_{\perp},z) &= 
\psi_{n_z^{\alpha}}(z) \Big(
\partial_{r_\perp} 
\psi_{n_{\perp}^{\alpha}}^{|\Lambda^{\! \alpha}|}(r_{\perp})\Big)
\psi_{n_z^{\beta}}(z) 
\psi_{n_{\perp}^{\beta}}^{|\Lambda^{\beta}|}(r_{\perp}) \nonumber \\
&- \psi_{n_z^{\alpha}}(z) 
\psi_{n_{\perp}^{\alpha}}^{|\Lambda^{\!\alpha}|}(r_{\perp})
\psi_{n_z^{\beta}}(z) \Big(
\partial_{r_\perp} 
\psi_{n_{\perp}^{\beta}}^{|\Lambda^{\beta}|}(r_{\perp})\Big), \\
\mathcal{F}^5_{\alpha \gamma}(r_{\perp},z) 
&= \frac{(\Lambda^{\!\alpha} \! + \! \Lambda^\beta)}{r_\perp} 
\mathcal{F}^1_{\alpha \gamma}(r_{\perp},z), \\
\mathcal{F}^6_{\alpha \gamma}(r_{\perp},z) &= 
\Big( \partial_z \psi_{n_z^{\alpha}}(z) \Big)
\psi_{n_{\perp}^{\alpha}}^{|\Lambda^{\! \alpha}|}(r_{\perp})
\psi_{n_z^{\beta}}(z) 
\psi_{n_{\perp}^{\beta}}^{|\Lambda^{\beta}|}(r_{\perp}) \nonumber \\
&- \psi_{n_z^{\alpha}}(z) 
\psi_{n_{\perp}^{\alpha}}^{|\Lambda^{\!\alpha}|}(r_{\perp})
\Big( \partial_z \psi_{n_z^{\beta}}(z) \Big) 
\psi_{n_{\perp}^{\beta}}^{|\Lambda^{\beta}|}(r_{\perp}),\\
\mathcal{F}^7_{\alpha \gamma}(r_{\perp},z) &= 
\frac{(\Lambda^{\!\alpha} \! - \! \Lambda^\beta)}{r_\perp}
\mathcal{F}^1_{\alpha \gamma}(r_{\perp},z).
\end{align}
\end{subequations}
Furthermore, the local pairing densities read
\begin{subequations}
\begin{align}
\tilde\rho^{(\tau)}(\bm{r})&= \sum_{\alpha \gamma} \kappa^{(\tau)}_{\alpha \gamma, -} \mathcal{F}^1_{\alpha \gamma}(r_{\perp},z)
\cos\Big[(\Lambda^{\!\alpha}\!-\!\Lambda^{\beta})\phi\Big], \\
\tilde\rho^{*(\tau)}(\bm{r})&= \sum_{\alpha \gamma} \kappa^{*(\tau)}_{\alpha \gamma, -} \mathcal{F}^1_{\alpha \gamma}(r_{\perp},z)
\cos\Big[(\Lambda^{\!\alpha}\!-\!\Lambda^{\beta})\phi\Big],
\end{align}
\end{subequations}
with an equivalent shorthand notation
\begin{subequations}
\begin{align}
\kappa_{\alpha \gamma, -}^{(\tau)} &= \frac{1}{2\pi}\Big( 
\kappa_{\alpha \gamma}^{(\tau) +-} \!-\! \kappa_{\alpha \gamma}^{(\tau) -+} \Big), \\
\kappa_{\alpha \gamma, -}^{*(\tau)} &= \frac{1}{2\pi}\Big( 
\kappa_{\alpha \gamma}^{*(\tau) +-} \!-\! \kappa_{\alpha \gamma}^{*(\tau) -+} \Big).
\end{align}
\end{subequations}

\section{Coupling Constants of the Skyrme EDF}
\label{sec:app_couplingconstants}
The time-even and time-odd contributions to the Skyrme
EDF [cf.~Eqs.~\eqref{eq:skyrme_even}
and \eqref{eq:skyrme_odd}, respectively] contain a total
of twenty coupling constants in the isoscalar ($t=0$)
and the isovector ($t=1$)
channel. Four of these constants
are density-dependent and can further be decomposed
as
\begin{subequations}
\begin{align}
C_{\bm{q},t}^{\rho \rho}(\bm{r};\bm{x}) &= C_{t,0}^{\rho \rho} + 
C_{t,D}^{\rho \rho} \rho_{\bm{q}}^{\alpha}(\bm{r};\bm{x}),\\
C_{\bm{q},t}^{s s}(\bm{r};\bm{x}) &= C_{t,0}^{s s} + 
C_{t,D}^{s s}\rho_{\bm{q}}^{\alpha}(\bm{r};\bm{x}).
\end{align}
\end{subequations}
Here, the real number $\alpha$ can be considered
as a parameter of an EDF. The remaining
twenty four
density-independent coupling constants 
can then be expressed
in terms of the $(t,x)$ parameters
of the Skyrme EDF. In the time-even
channel, the coupling constants read
\begin{subequations}
\begin{align}
C_{0,0}^{\rho \rho} &= + \frac{3}{8}t_0, \\
C_{0,D}^{\rho \rho} &= + \frac{1}{16}t_3, \\
C_{1,0}^{\rho \rho} &= -\frac{1}{4}t_0\Big(\frac{1}{2}+x_0\Big), \\
C_{1,D}^{\rho \rho} &= -\frac{1}{24}t_3\Big(\frac{1}{2}+x_3\Big), \\
C_0^{\rho \Delta \rho} &= - \frac{9}{64}t_1 
+  \frac{1}{16}t_2\Big(\frac{5}{4}+x_2\Big), \\
C_1^{\rho \Delta \rho} &= +\frac{3}{32}t_1
\Big(\frac{1}{2}+x_1\Big) +  \frac{1}{32}t_2\Big(\frac{1}{2}+x_2\Big), \\
C_0^{\rho \tau} &= +\frac{3}{16}t_1 + \frac{1}{4}t_2\Big(\frac{5}{4}+x_2\Big), \\
C_1^{\rho \tau} &= -\frac{1}{8}t_1\Big(\frac{1}{2}+x_1\Big) + \frac{1}{8}t_2\Big(\frac{1}{2}+x_2\Big),\\
C_0^{\rho \nabla J} &= -b_4 - \frac{1}{2}b_4', \\ 
C_1^{\rho \nabla J} &= - \frac{1}{2}b_4', \\
C_0^{JJ} &= +\frac{1}{8}t_1\Big(\frac{1}{2}-x_1\Big) - \frac{1}{8}t_2\Big(\frac{1}{2}+x_2\Big), \\
C_1^{JJ} &= -\frac{1}{16}\Big(t_2-t_1\Big),
\end{align}
\end{subequations}
where $b_4$ and $b_4'$ are the parameters 
of the spin-orbit force
and we took $t_e = t_o = 0$ for
the tensor terms \cite{schunck2019energy}.
In the time-odd
channel, the coupling constants read
\begin{subequations}
\begin{align}
C_{0,0}^{s s } &= - \frac{1}{4}t_0\Big(\frac{1}{2}-x_0\Big), \\
C_{0,D}^{s s} &= - \frac{1}{24}t_3\Big(\frac{1}{2}-x_3\Big), \\
C_{1,0}^{s s} &= -\frac{1}{8}t_0, \\
C_{1,D}^{s s} &= - \frac{1}{48}t_3, \\
C_0^{s\Delta s} &= + \frac{3}{32}t_1 
(\frac{1}{2}-x_1\Big)+  \frac{1}{32}t_2\Big(\frac{1}{2}+x_2\Big), \\
C_1^{s \Delta s} &= +\frac{3}{64}t_1
+  \frac{1}{64}t_2, \\
C_0^{sj} &=-C_0^{\rho \tau}, \label{eq:csj0} \\
C_1^{sj} &= -C_1^{\rho \tau},\\
C_0^{s \nabla j} &= +C_0^{\rho \nabla J}, \\ 
C_1^{s \nabla j} &= +C_1^{\rho \nabla J}, \\
C_0^{sT} &= -C_0^{JJ}, \\
C_1^{sT} &= -C_1^{JJ} \label{eq:csT1}.
\end{align}
\end{subequations}
Note that relations \eqref{eq:csj0}~-~\eqref{eq:csT1} are 
imposed by the local gauge invariance of 
an EDF \cite{schunck2019energy}.

\bibliographystyle{elsarticle-num}
\bibliography{zotero_output,books}

\newcommand{\noopsort}[1]{}
\begin{thebibliography}{10}
\expandafter\ifx\csname url\endcsname\relax
  \def\url#1{\texttt{#1}}\fi
\expandafter\ifx\csname urlprefix\endcsname\relax\def\urlprefix{URL }\fi
\expandafter\ifx\csname href\endcsname\relax
  \def\href#1#2{#2} \def\path#1{#1}\fi

\bibitem{schunck2019energy}
N.~Schunck, Energy Density Functional Methods for Atomic Nuclei., {{IOP
  Expanding Physics}}, {IOP Publishing}, {Bristol, UK}, 2019.

\bibitem{bender2003selfconsistent}
M.~Bender, P.-H. Heenen, P.-G. Reinhard, Self-consistent mean-field models for
  nuclear structure, Rev. Mod. Phys. 75~(1) (2003) 121.

\bibitem{niksic2011relativistic}
T.~Nik{\v s}i{\'c}, D.~Vretenar, P.~Ring, Relativistic nuclear energy density
  functionals: {{Mean}}-field and beyond, Prog. Part. Nucl. Phys. 66~(3) (2011)
  519.

\bibitem{robledo2019mean}
L.~M. Robledo, T.~R. Rodr{\'i}guez, R.~R. {Rodr{\'i}guez-Guzm{\'a}n}, Mean
  field and beyond description of nuclear structure with the {{Gogny}} force: A
  review, J. Phys. G: Nucl. Part. Phys. 46~(1) (2019) 013001.

\bibitem{perez2017axially}
R.~N. Perez, N.~Schunck, R.-D. Lasseri, C.~Zhang, J.~Sarich, Axially deformed
  solution of the {{Skyrme}}--{{Hartree}}--{{Fock}}--{{Bogolyubov}} equations
  using the transformed harmonic oscillator basis ({{III}}) {{HFBTHO}} (v3.00):
  {{A}} new version of the program, Comput. Phys. Commun. 220 (2017) 363.

\bibitem{schunck2017solution}
N.~Schunck, J.~Dobaczewski, W.~Satu{\l}a, P.~B{\k{a}}czyk, J.~Dudek, Y.~Gao,
  M.~Konieczka, K.~Sato, Y.~Shi, X.~B. Wang, T.~R. Werner, Solution of the
  {{Skyrme}}-{{Hartree}}-{{Fock}}-{{Bogolyubov}} equations in the {{Cartesian}}
  deformed harmonic-oscillator basis. ({{VIII}}) {{HFODD}} (v2.73y): {{A}} new
  version of the program, Comput. Phys. Commun. 216 (2017) 145.

\bibitem{ryssens2015solution}
W.~Ryssens, V.~Hellemans, M.~Bender, P.-H. Heenen, Solution of the
  {{Skyrme}}-{{HF}}+{{BCS}} equation on a {{3D}} mesh, {{II}}: {{A}} new
  version of the {{Ev8}} code, Comput. Phys. Commun. 187 (2015) 175.

\bibitem{niksic2014dirhb}
T.~Nik{\v{s}}i{\'c}, N.~Paar, D.~Vretenar, P.~Ring, {{DIRHB}} - {{A}}
  relativistic self-consistent mean-field framework for atomic nuclei, Comput.
  Phys. Commun. 185~(6) (2014) 1808.

\bibitem{carlsson2010solution}
B.~G. Carlsson, J.~Dobaczewski, J.~Toivanen, P.~Vesel{\'y}, Solution of
  self-consistent equations for the {{N3LO}} nuclear energy density functional
  in spherical symmetry. {{The}} program {{HOSPHE}} (v1.02), Comput. Phys.
  Commun. 181~(9) (2010) 1641.

\bibitem{bennaceur2005coordinatespace}
K.~Bennaceur, J.~Dobaczewski, Coordinate-space solution of the
  {{Skyrme}}-{{Hartree}}-{{Fock}}- {{Bogolyubov}} equations within spherical
  symmetry. {{The}} program {{HFBRAD}} (v1.00), Comput. Phys. Commun. 168~(2)
  (2005) 96.

\bibitem{stoitsov2005axially}
M.~V. Stoitsov, J.~Dobaczewski, W.~Nazarewicz, P.~Ring, Axially deformed
  solution of the {{Skyrme}}-{{Hartree}}-{{Fock}}-{{Bogolyubov}} equations
  using the transformed harmonic oscillator basis. {{The}} program {{HFBTHO}}
  (v1.66p), Comput. Phys. Commun. 167~(1) (2005) 43.

\bibitem{stoitsov2013axially}
M.~Stoitsov, N.~Schunck, M.~Kortelainen, N.~Michel, H.~Nam, E.~Olsen,
  J.~Sarich, S.~Wild, Axially deformed solution of the
  {{Skyrme}}-{{Hartree}}-{{Fock}}-{{Bogoliubov}} equations using the
  transformed harmonic oscillator basis ({{II}}) {{HFBTHO}} v2.00d: {{A}} new
  version of the program, Comput. Phys. Commun. 184~(6) (2013) 1592.

\bibitem{dobaczewski2021solution}
J.~Dobaczewski, P.~B{\k{a}}czyk, P.~Becker, M.~Bender, K.~Bennaceur,
  J.~Bonnard, Y.~Gao, A.~Idini, M.~Konieczka, M.~Kortelainen, L.~Pr{\'o}chniak,
  A.~M. Romero, W.~Satu{\textbackslash}la, Y.~Shi, T.~R. Werner, L.~F. Yu,
  Solution of universal nonrelativistic nuclear {{DFT}} equations in the
  {{Cartesian}} deformed harmonic-oscillator basis. ({{IX}}) {{HFODD}}
  (v3.06h): A new version of the program, J. Phys. G: Nucl. Part. Phys. 48~(10)
  (2021) 102001.

\bibitem{sheikh2021symmetry}
J.~A. Sheikh, J.~Dobaczewski, P.~Ring, L.~M. Robledo, C.~Yannouleas, Symmetry
  restoration in mean-field approaches, J. Phys. G: Nucl. Part. Phys. 48~(12)
  (2021) 123001.

\bibitem{bally2021projection}
B.~Bally, M.~Bender, Projection on particle number and angular momentum:
  {{Example}} of triaxial {{Bogoliubov}} quasiparticle states, Phys. Rev. C
  103~(2) (2021) 024315.

\bibitem{varshalovich1988}
D.~Varshalovich, A.~Moskalev, V.~Khersonskii, Quantum Theory of Angular
  Momentum, World Scientific, Singapore, 1988.

\bibitem{ring2004}
P.~Ring, P.~Schuck, The Nuclear Many-Body Problem, Texts and Monographs in
  Physics, Springer, 2004.

\bibitem{fomenko1970projection}
V.~N. Fomenko, Projection in the occupation-number space and the canonical
  transformation, J. Phys. A: Gen. Phys. 3~(1) (1970) 8.

\bibitem{egido1991parityprojected}
J.~L. Egido, L.~M. Robledo, Parity-projected calculations on octupole deformed
  nuclei, Nucl. Phys. A 524~(1) (1991) 65.

\bibitem{dobaczewski2000point}
J.~Dobaczewski, J.~Dudek, S.~G. Rohozi{\'n}ski, T.~R. Werner, Point symmetries
  in the {{Hartree}}-{{Fock}} approach. {{I}}. {{Densities}}, shapes, and
  currents, Phys. Rev. C 62~(1) (2000) 014310.

\bibitem{dobaczewski2000pointa}
J.~Dobaczewski, J.~Dudek, S.~G. Rohozi{\'n}ski, T.~R. Werner, Point symmetries
  in the {{Hartree}}-{{Fock}} approach. {{II}}. {{Symmetry}}-breaking schemes,
  Phys. Rev. C 62~(1) (2000) 014311.

\bibitem{egido2016stateoftheart}
J.~L. Egido, State-of-the-art of beyond mean field theories with nuclear
  density functionals, Phys. Scr. 91~(7) (2016) 073003.

\bibitem{avez2012evaluation}
B.~Avez, M.~Bender, Evaluation of overlaps between arbitrary fermionic
  quasiparticle vacua, Phys. Rev. C 85~(3) (2012) 034325.

\bibitem{valor2000configuration}
A.~Valor, P.~H. Heenen, P.~Bonche, Configuration mixing of mean-field wave
  functions projected on angular momentum and particle number: {{Application}}
  to {{$^{24}$Mg}}, Nucl. Phys. A 671~(1) (2000) 145.

\bibitem{baye1984angular}
D.~Baye, P.-H. Heenen, Angular momentum projection on a mesh of cranked
  {{Hartree}}-{{Fock}} wave functions, Phys. Rev. C 29~(3) (1984) 1056.

\bibitem{robledo1994practical}
L.~M. Robledo, Practical formulation of the extended {{Wick}}'s theorem and the
  {{Onishi}} formula, Phys. Rev. C 50~(6) (1994) 2874.

\bibitem{balian1969nonunitary}
R.~Balian, E.~Brezin, Nonunitary bogoliubov transformations and extension of
  {{Wick}}'s theorem, Nuovo Cim. B 64~(1) (1969) 37.

\bibitem{hara1979quantum}
K.~Hara, S.~Iwasaki, On the quantum number projection, Nucl. Phys. A 332~(1)
  (1979) 61.

\bibitem{marevic2020fission}
P.~Marevi{\'c}, N.~Schunck, Fission of $^{240}\mathrm{Pu}$ with
  {{Symmetry}}-{{Restored Density Functional Theory}}, Phys. Rev. Lett.
  125~(10) (2020) 102504.

\bibitem{marevic2021angular}
P.~Marevi{\'c}, N.~Schunck, J.~Randrup, R.~Vogt, Angular momentum of fission
  fragments from microscopic theory, Phys. Rev. C 104~(2) (2021) L021601.

\bibitem{onishi1966generator}
N.~Onishi, S.~Yoshida, Generator coordinate method applied to nuclei in the
  transition region, Nucl. Phys. 80~(2) (1966) 367.

\bibitem{nazmitdinov1996representation}
R.~G. Nazmitdinov, L.~M. Robledo, P.~Ring, J.~L. Egido, Representation of
  three-dimensional rotations in oscillator basis sets, Nucl. Phys. A 596~(1)
  (1996) 53.

\bibitem{rohozinski2010selfconsistent}
S.~G. Rohozi{\'n}ski, J.~Dobaczewski, W.~Nazarewicz, Self-consistent symmetries
  in the proton-neutron {{Hartree}}-{{Fock}}-{{Bogoliubov}} approach, Phys.
  Rev. C 81~(1) (2010) 014313.

\bibitem{engel1975timedependent}
Y.~M. Engel, D.~M. Brink, K.~Goeke, S.~J. Krieger, D.~Vautherin, Time-dependent
  {{Hartree}}-{{Fock}} theory with {{Skyrme}}'s interaction, Nucl. Phys. A
  249~(2) (1975) 215.

\bibitem{perez-martin2008microscopic}
S.~{Perez-Martin}, L.~Robledo, Microscopic justification of the equal filling
  approximation, Phys. Rev. C 78~(1) (2008) 014304.
\newblock \href {https://doi.org/10.1103/PhysRevC.78.014304}
  {\path{doi:10.1103/PhysRevC.78.014304}}.

\bibitem{robledo2007particle}
L.~M. Robledo, Particle number restoration: Its implementation and impact in
  nuclear structure calculations, Int. J. Mod. Phys. E 16~(02) (2007) 337--351.

\bibitem{robledo2010remarks}
L.~M. Robledo, Remarks on the use of projected densities in the
  density-dependent part of {{Skyrme}} or {{Gogny}} functionals, J. Phys. G:
  Nucl. Part. Phys. 37~(6) (2010) 064020.

\bibitem{bulgac2018minimal}
A.~Bulgac, M.~M. Forbes, S.~Jin, R.~N. Perez, N.~Schunck, Minimal nuclear
  energy density functional, Phys. Rev. C 97~(4) (2018) 044313.

\bibitem{dobaczewski2002contact}
J.~Dobaczewski, W.~Nazarewicz, M.~V. Stoitsov, Contact pairing interaction for
  the {{Hartree}}-{{Fock}}-{{Bogoliubov}} calculations, in: The {{Nuclear
  Many}}-{{Body Problem}} 2001, no.~53 in Nato {{Science Series II}}, {Springer
  Netherlands}, 2002, p. 181.

\bibitem{bulgac2002local}
A.~Bulgac, Local density approximation for systems with pairing correlations,
  Phys. Rev. C 65~(5) (2002) 051305(R).

\bibitem{girod1983triaxial}
M.~Girod, B.~Grammaticos, Triaxial {{Hartree}}-{{Fock}}-{{Bogolyubov}}
  calculations with {{D1}} effective interaction, Phys. Rev. C 27~(5) (1983)
  2317.

\bibitem{dobaczewski2009solution}
J.~Dobaczewski, W.~Satu{\l}a, B.~G. Carlsson, J.~Engel, P.~Olbratowski,
  P.~Powa{\l}owski, M.~Sadziak, J.~Sarich, N.~Schunck, A.~Staszczak,
  M.~Stoitsov, M.~Zalewski, H.~Zdu{\'n}czuk, Solution of the
  {{Skyrme}}-{{Hartree}}-{{Fock}}-{{Bogolyubov}} equations in the {{Cartesian}}
  deformed harmonic-oscillator basis. ({{VI}}) {{HFODD}} (v.240h): {{A}} new
  version of the program, Comput. Phys. Commun. 180~(11) (2009) 2361.

\bibitem{dobaczewski1996meanfield}
J.~Dobaczewski, W.~Nazarewicz, T.~R. Werner, J.~F. Berger, C.~R. Chinn,
  J.~Decharg{\'e}, Mean-field description of ground-state properties of
  drip-line nuclei: {{Pairing}} and continuum effects, Phys. Rev. C 53~(6)
  (1996) 2809.
\newblock \href {https://doi.org/10.1103/PhysRevC.53.2809}
  {\path{doi:10.1103/PhysRevC.53.2809}}.

\bibitem{beiner1975nuclear}
M.~Beiner, H.~Flocard, N.~Van~Giai, P.~Quentin, Nuclear ground-state properties
  and self-consistent calculations with the skyrme interaction:({{I}}).
  {{Spherical}} description, Nucl. Phys. A 238~(1) (1975) 29.

\bibitem{chabanat1998skyrme}
E.~Chabanat, P.~Bonche, P.~Haensel, J.~Meyer, R.~Schaeffer, A {{Skyrme}}
  parametrization from subnuclear to neutron star densities {{Part II}}.
  {{Nuclei}} far from stabilities, Nucl. Phys. A 635~(1) (1998) 231.

\bibitem{reinhard1999shape}
P.-G. Reinhard, D.~J. Dean, W.~Nazarewicz, J.~Dobaczewski, J.~A. Maruhn, M.~R.
  Strayer, Shape coexistence and the effective nucleon-nucleon interaction,
  Phys. Rev. C 60~(1) (1999) 014316.

\bibitem{valor1997new}
A.~Valor, J.~L. Egido, L.~M. Robledo, A new approach to approximate symmetry
  restoration with density dependent forces: {{The}} superdeformed band in
  {{192Hg}}, Phys. Lett. B 392~(3) (1997) 249.

\bibitem{schunck2013density}
N.~Schunck, Density {{Functional Theory Approach}} to {{Nuclear Fission}}, Acta
  Phys. Pol. B 44 (2013) 263.

\bibitem{anguiano2001particle}
M.~Anguiano, J.~L. Egido, L.~M. Robledo, Particle number projection with
  effective forces, Nucl. Phys. A 696~(3\textendash 4) (2001) 467.

\bibitem{dobaczewski2007particlenumber}
J.~Dobaczewski, M.~V. Stoitsov, W.~Nazarewicz, P.-G. Reinhard, Particle-number
  projection and the density functional theory, Phys. Rev. C 76~(5) (2007)
  054315.

\bibitem{duguet2009particlenumber}
T.~Duguet, M.~Bender, K.~Bennaceur, D.~Lacroix, T.~Lesinski, Particle-number
  restoration within the energy density functional formalism: {{Nonviability}}
  of terms depending on noninteger powers of the density matrices, Phys. Rev. C
  79~(4) (2009) 044320.

\bibitem{bender2009particlenumber}
M.~Bender, T.~Duguet, D.~Lacroix, Particle-number restoration within the energy
  density functional formalism, Phys. Rev. C 79~(4) (2009) 044319.

\bibitem{schunck2008continuum}
N.~Schunck, J.~L. Egido, Continuum and symmetry-conserving effects in drip-line
  nuclei using finite-range forces, Phys. Rev. C 77~(1) (2008) 011301(R).

\bibitem{schunck2008nuclear}
N.~Schunck, J.~L. Egido, Nuclear halos and drip lines in symmetry-conserving
  continuum {{Hartree}}-{{Fock}}-{{Bogoliubov}} theory, Phys. Rev. C 78~(6)
  (2008) 064305.

\bibitem{ney2020global}
E.~M. Ney, J.~Engel, T.~Li, N.~Schunck, Global description of $\beta^{-}$ decay
  with the axially deformed {{Skyrme}} finite-amplitude method: {{Extension}}
  to odd-mass and odd-odd nuclei, Phys. Rev. C 102~(3) (2020) 034326.

\end{thebibliography}


\begin{thebibliography}{0}
\bibitem{stoitsov2005}
M.~V. Stoitsov, J.~Dobaczewski, W.~Nazarewicz, P.~Ring, Axially deformed
  solution of the {{Skyrme}}-{{Hartree}}-{{Fock}}-{{Bogolyubov}} equations
  using the transformed harmonic oscillator basis. {{The}} program {{{\hfbtho}}}
  (v1.66p), Comput. Phys. Commun. 167~(1) (2005) 43.
\bibitem{stoitsov2013}
M.~Stoitsov, N.~Schunck, M.~Kortelainen, N.~Michel, H.~Nam, E.~Olsen,
  J.~Sarich, S.~Wild, Axially deformed solution of the
  {{Skyrme}}-{{Hartree}}-{{Fock}}-{{Bogolyubov}} equations using the
  transformed harmonic oscillator basis ({{II}}) {{{\hfbtho}}} v2.00d: {{A}} new
  version of the program, Comput. Phys. Commun. 184~(6) (2013) 1592.
\bibitem{perez2017}
R.~N. Perez, N.~Schunck, R.-D. Lasseri, C.~Zhang, J.~Sarich, Axially deformed
  solution of the {{Skyrme}}--{{Hartree}}--{{Fock}}--{{Bogolyubov}} equations
  using the transformed harmonic oscillator basis ({{III}}) {{{\hfbtho}}} (v3.00):
  {{A}} new version of the program, Comput. Phys. Commun. 220~
  (2017) 363.
\end{thebibliography}

\end{document}